\documentclass[12pt]{article}

\usepackage{latexsym,amsmath,amsfonts,amssymb}

\usepackage{epsfig}

\renewcommand{\theequation}{\arabic{section}.\arabic{equation}}
\newcommand{\newsection}{    
\setcounter{equation}{0}\section}

\newcommand{\acknowledgments} {\section*{Acknowledgments}}

\def\be{\begin{equation}}
\def\ee{\end{equation}}
\newcommand{\bea}{\begin{eqnarray}}
\newcommand{\eea}{\end{eqnarray}}
\def\bd{\begin{displaymath}}
\def\ed{\end{displaymath}}
\def\alppr{{\alpha^\prime}}

\def\ba{\begin{eqnarray}}
\def\ea{\end{eqnarray}}

\allowdisplaybreaks[3]

\begin{document}

\vspace{18pt}
\par\hfill CPHT-RR-059.1005
\vskip 0.01in \hfill TAUP-2814/05
\vskip 0.01in \hfill hep-th/0510110

\vspace{30pt}

\begin{center}
{\bf \LARGE Fundamental matter, meson spectroscopy \\[5pt] and non-critical string/gauge duality}
\end{center}

\vspace{30pt}

\begin{center}
Roberto Casero$\,^a$, Angel Paredes$\,^a$, Jacob Sonnenschein$\,^b$

\vspace{30pt}

\textit{$^a$Centre de Physique Th\'eorique\\ \'Ecole Polytechnique\\ 91128 Palaiseau, France\\[15pt]
$^b$School of Physics and Astronomy\\ The Raymond and Beverly Sackler Faculty of Exact Sciences\\ Tel Aviv University\\ Ramat Aviv 69978, Israel}

\end{center}


\begin{center}
\textbf{Abstract }
\end{center}

\vspace{4pt} {\small \noindent 
We discuss the incorporation of quarks in the fundamental representation 
of the color group into the non-critical string/gauge duality.
We focus on confining theories and address this question using two 
different approaches: (i) by 
introducing flavor probe branes and (ii) by deriving backreacted 
flavored near extremal
gravity backgrounds. In the former  approach  we analyze the
near extremal $AdS_6$ model  with  D4 and anti-D4 
probe flavor branes included.
We study the meson spectrum and discuss the role played by the
constituent quark mass, related to the integration constant that defines
the embedding.
As for the second approach we derive a class of flavored 
$AdS_{n+1}\times S^k$ black hole solutions. In particular we write down  the 
flavored   $AdS_6$ and $AdS_5$ black holes and the near 
extremal $AdS_5\times S^1$ 
backgrounds.
We analyze several gauge dynamical properties associated with these 
models.
 }  \vfill
\vskip 5.mm
 \hrule width 5.cm
\vskip 2.mm
{\small
\noindent
roberto.casero@cpht.polytechnique.fr\\ angel.paredes@cpht.polytechnique.fr\\ cobi@post.tau.ac.il}

\thispagestyle{empty}

\eject

\setcounter{page}{1}

\newsection{Introduction}
The program of constructing the string theory holographic dual to QCD has made a significant progress in recent years. 
However, there are still several major open questions  that have to be addressed. These include in particular the following
two challenges:  
 \begin{itemize}
\item
 A full understanding of the consistent incorporation of flavored quarks in the fundamental 
representation of the color group; 
\item
 Cleaning the spectrum
from the undesired KK modes associated with the compact  transverse directions.
\end{itemize}
In this paper we address these two issues in the context of non-critical 
(spacetime dimension less than ten) string/gauge  dualities
\cite{Polyakov}-\cite{BCCKP}.

The identification of the particles  of confining gauge theories in their dual supergravity (string) 
 models can be achieved  using the following  three different approaches:
\begin{enumerate}
\item 
 For gravity backgrounds that incorporate bulk fields which  correspond to the assigned states of the boundary gauge theory,
one computes the spectrum of the fluctuations of those bulk fields and deduces from it the spectrum of the dual particles.
\item
For states of the gauge theory  that do not have associated  bulk  fields in the dual gravity  theory,
one introduces probe branes  in such a way that their    fluctuations play the role of the  desired states. 
\item
 For states of spin higher than two one has to go beyond the  gravity   limit and consider string configurations that describe the particles. 
\end{enumerate}
A prototype of the  
 identification of the first kind is the four-dimensional 
glueball spectrum extracted from the fluctuations of the dilaton (and other fields)
in  confining backgrounds~\cite{oz}, such as the model based on  near extremal D4 branes.

To dualize the   fundamental quarks one can either derive a string theory or its gravity low-energy effective action
that incorporates fields dual to the quarks or else  use approach~2 of introducing flavor probe branes. 
Since most of the known supergravity backgrounds do not incorporate fields dual to fundamental quarks, it was the idea of
\cite{Karch:2003nh} to introduce probe branes in such a way that  strings stretching between them and the original branes have the properties 
of flavored fundamental quarks. In the original model \cite{Karch:2003nh} $N_f$ D7 probe  branes were incorporated in the $AdS_5\times S^5$ supergravity 
and the strings between the D3 and D7 branes were the duals of  hyper-multiplets of the 
resulting N=2 SQCD theory in the fundamental representation of both the color and flavor groups.
The fluctuations of the probe branes were shown \cite{KMMW} to correspond  to  the mesons of this theory which 
is  in a Coulomb phase.
 The first attempt to describe mesons of a confining gauge theory was made 
 in \cite{SakSon} where D7 probe branes were introduced into the
 Klebanov-Strassler model \cite{KS},
 and  the spectrum of massive vector and pseudoscalar mesons was computed.
A much simpler and more illuminating model  was proposed in \cite {KMMW2} 
which is based on the addition of  D6 flavor branes to the model  of near extremal 
D4 branes \cite{Wpure}.  
For further work devoted to the addition of flavors
in the context of critical string duality, see, for instance,
\cite{problist}-\cite{Nunez:2003cf}.
Related to this context, several holographic models for hadrons have also been built, where an  {\it ad hoc}  IR cutoff is introduced in $AdS_5$  in
order to make the dual theory confining and with a mass gap \cite{mesonlist} (see also
\cite{Mahajan:2005rk}), achieving notable success for their quantitative 
predictions. However, it is of obvious interest to obtain this kind of
five-dimensional actions from string theory.

Recently, progress in this direction was made by incorporating D8,$\rm{\bar{D8}}$
 branes in the near extremal D4 brane background \cite{Wpure}, resulting in a model 
with a non-abelian flavor  symmetry and flavor chiral symmetry breaking \cite{SS}.
In particular the corresponding Goldstone bosons were identified.
Furthermore,  ratios of the masses of certain vector and  scalar 
mesons were computed  in this model 
and were compared  to the  experimental values  yielding a fairly nice agreement. In a later work, the same  authors computed in their model decay rates of certain mesons \cite{SS2}.
All these models are examples of the second type of approach mentioned above.

Glueballs and mesons of spin higher than two cannot be described by a  low
energy gravitational theory simply because
 the latter does not include fields of spin higher than two. The hadronic spectrum of confining gauge theories  however does include hadrons of higher spin and  the lightest ones for each spin furnish a Regge trajectory, namely, they obey the relation  $J=\alpha' M^2 + \alpha_0$. 
It is well known that this kind of relation  between energy and  angular momentum is the characteristic behavior of open strings in flat spacetime. Recently, it has been  shown that string configurations that reside in the vicinity of the ``wall'' of confining backgrounds also admit the Regge trajectory behavior.
In particular in \cite{PSV} using a semi-classical quantization 
this  was shown for  closed strings, the duals of glueballs, in the context of
the Klebanov-Strassler \cite{KS} and Maldacena-N\'u\~nez~\cite{MN} models, and in \cite{Leo} for the near extremal 
D4 branes background.
Regarding high spin mesons, it was shown in \cite{KPSV} that the open strings that describe them are equivalent to 
open strings with massive endpoints spinning in flat space.   

These developments address the first challenge mentioned above. 
As for the second, the most naive approach to the problem of the KK modes 
is  the use of non-critical string theories where the KK (or at least most of
them) are absent a priori.
The idea to employ a non-critical string theory as the string theory of QCD was
raised in \cite{Polyakov}.
This proposal was further investigated in several other papers \cite{noncrihol}. Recently,  a few families of 
non-critical backgrounds were identified and analyzed \cite{KS1}, including  in particular a class of models with spacetimes of the form
$AdS_{n+1}\times S^k$, a constant dilaton and an $F_{n+1}$ RR form. 
``Measurements'' of various properties were performed in the $AdS_6$ black hole ``laboratory'' in \cite{KS2}.

Non-critical supergravities suffer from the fact
that they have necessarily finite curvature in units of the 
string scale due to the non-critical term in their action.
However, there is some evidence that the basic structure of 
spacetime is not modified  due to higher order curvature corrections
in a similar manner to what occurs for strings on group manifolds.  
In this paper we follow this approach and consider non-critical 
gravity 
models. 
On the other hand, recent developments using a worldsheet approach to non-critical
strings can be found in \cite{noncrit}.

To introduce fundamental quarks into  non-critical models, one can  
 again follow one of the  two alternatives above: to derive a fully backreacted 
  gravity background or to
analyze a system with probe branes.     In the present work
we discuss both options.

The  ``laboratory'' to which we add flavor probe branes is the $AdS_6$ non-critical black hole solution \cite{KS1}. This model  was shown \cite{KS2} to reproduce some
properties of 4d non-supersymmetric YM theory like an area law for the Wilson loop, a mass gap in the glueball spectrum etc.
In this work we introduce flavor in this setup by adding D4,$\rm{\bar{D4}}$ 
probe branes. This is similar to the D8,$\rm{\bar{D8}}$ model of \cite{SS}
in critical dimension and it inherits from it many qualitative features.
We derive the profile of the probe branes deduced from an action with or without a CS term. 
We analyze the pseudoscalar and vector fluctuations on the worldvolume of the probe branes
and extract the spectrum of the corresponding pseudoscalar and vector mesons. 
Similarly to \cite{SS} we find  massless Goldstone bosons. We generalize the 
solutions of \cite{SS} by allowing the lowest point of the 
probe brane to be away from  the ``wall" of the spacetime. 
We assign to the difference between the former and the latter 
a ``mass parameter'' $m_q$. 
We find that the masses of the low spin mesons increase 
monotonically with  this   mass parameter (for a probe action that 
includes a CS term this increase takes place only after 
a certain minimal value of $m_q$). 
We compute the ratios of  the masses of  various mesons and find that  the  
agreement with the measured values  is of the same order as  
those of the critical model~\cite{SS}. 
 It turns out, however, that even for a non-trivial mass parameter 
 the  ``pions'' remain massless. This result indicates that,
in the terminology
of QCD, $m_q$ does not correspond to the 
current algebra mass of the quarks (see section \ref{sect: mass}
for a discussion).
We then analyze the spectrum of classical spinning open string 
configurations and find that as for the 
critical case, one can describe the mesons as  spinning open 
strings in flat spacetime with massive endpoints whose mass is $m_q$.

As for the fully backreacted effective gravitational actions, 
some solutions have been computed in critical string theory \cite{cherkis}.
In several non-critical models \cite{KM,alish,BCCKP},
 open string degrees of freedom that are dual to  
 quarks in the 
 color and flavor fundamental representation have
 been incorporated by adding a DBI term to the gravity action.
The Klebanov-Maldacena (KM) \cite{KM}  model is an $AdS_5\times S^1$
 supergravity background which 
 is conjectured to be the holographic dual of  the IR fixed point
of the  $N=1$ SQCD  in the so called ``conformal window''. In \cite{BCCKP} two other flavored models were derived in 5d and 8d corresponding to a QCD-like theory with $N=0$ and $N=1$ supersymmetries respectively.

A natural question to address at this stage is the derivation of  
a  background which is dual to a  flavored gauge theory which is not
conformal but rather confining, 
 for instance  non-supersymmetric QCD. One can address this challenge 
 by either starting with 
unflavored non-conformal backgrounds and add  quarks to them, or by taking  a conformal flavored model, break
conformal invariance and turn on confinement. 
In this paper we do both. We derive non-critical non-conformal flavored backgrounds by 
starting with the non-conformal
near extremal $AdS_6$ ($AdS_5$) model, which describes non-critical near extremal 
D4 (D3)  branes. We then 
introduce $N_f$ flavor branes and write down the system of
equations that describes the fully backreacted 
solution  of the gravity equations of motion.
 In the  $AdS_6$ case, with D4 flavor branes smeared over the entire 
``thermal circle''
the only solution we could determine explicitly 
is eventually a conformal solution, the T-dual of the KM 
model.
On the other hand for space-filling D5 branes  we do find a 
 flavored $AdS_6$ black hole solution.
As in the original proposal of \cite{Wpure} upon turning on large 
temperature or reducing the radius of the thermal circle we end up with a 
four-dimensional low-energy effective action. Unfortunately, this theory 
includes only the pure YM theory since the 
flavored fundamental quarks also acquire mass of the order of the 
temperature and decouple from the low-energy effective theory.  
We then generalize the KM model and derive a family of flavored 
conformal supergravity solutions in $D$ dimensions of the form 
 $AdS_{n+1}\times S^k$, 
by incorporating $N_f$ space-filling DBI terms into the gravity action. 
We also derive the corresponding black  
hole solutions, and in particular the near extremal limit of the KM model  
 $AdS_5\times S^1$ \cite{KM} . 

We address the issue of the 
 ``phenomenology'' of the flavored near extremal models that are dual to
 four-dimensional gauge systems.
We present several arguments about the identification of the various 
phases of the latter and then
extract gauge dynamical properties such as
 the Wilson loop, the 
spectrum of glueballs and the spectrum of mesons.

The paper is organized as follows: after this introduction, in section 
\ref{sect: AdS6}
the general setup of the near extremal 
$AdS_6$  non-critical background is described. We then discuss 
qualitatively the options to introduce flavor branes into this model. 
Section 
\ref{probe D4} is devoted to a quantitative analysis of the system when D4 
probe branes are incorporated.
The solution for the classical probe profile is written down and  the spectrum 
of the fluctuations of the gauge fields as well as the one of (pseudo) 
scalar fields is extracted. This includes the identification of the 
Goldstone bosons, the computation of the masses of the low-lying mesons 
and a discussion of the dependence of the these features on the CS 
coefficient
and on $m_q$, the mass parameter of the probe profile. We then analyze 
the configurations of open spinning strings.
It is shown that similarly to the result reported in \cite{KPSV}, these 
configurations behave like open strings in flat spacetime with massive 
endpoints. 
In section \ref{flavor D4}
 we attempt at finding  a backreacted solution of $N_f$ 
flavor D4 branes that are smeared along the thermal cycle. The only
constant dilaton solution the system admits 
is an $AdS_5\times S^1$ geometry which is T-dual to the KM solution.
We devote section \ref{sect: general} to the study of 
solutions associated with space-filling probe branes in any number $D$ of 
dimensions and with an $S^k$ transverse space. Both conformal solutions and their 
black hole generalizations are derived. 
In sections \ref{flavor D5}-\ref{Ads5} we present separately the 
most interesting cases.
In section 
\ref{sect: phenom} the phenomenology of the 
gauge theories dual to some of these backgrounds is extracted.
We end up with a 
summary and a list of several open questions in section \ref{summary}.


\newsection{Towards a non-critical string dual of QCD}
\label{sect: AdS6}

The six-dimensional background we consider in this section is the non-critical version \cite{KS1} of the ten-dimensional black hole background dual to thermal/pure Yang-Mills in four dimensions \cite{Wpure}. In both the critical and non-critical cases, the gravity background is generated by near-extremal D4-branes wrapped over a circle with anti-periodic boundary conditions.
In the non-critical case it takes the form of  a static black hole embedded inside six-dimensional anti-de Sitter space. The anti-periodic boundary conditions project massless fermions out of the spectrum, and the only surviving fermionic degrees of freedom are excited KK modes. In turn, this also gives a mass to all the scalars via one-loop corrections. At high energies, therefore,  the D4-brane dynamics describes a five-dimensional gauge theory at the same finite temperature as the black hole temperature, whereas at low-energies, the KK modes cannot be excited and the theory is effectively pure Yang-Mills in four dimensions \cite{Wpure}.

The background geometry
 consists of a constant dilaton $\phi$, a RR six-form field strength $F_{(6)}$
and a metric \cite{KS1}:
\bea
ds^2_6&=&\left( \frac{u}{R_{AdS}} \right)^2 dx_{1,3}^2
+\left( \frac{R_{AdS}}{u} \right)^2 \frac{du^2}{f(u)} +
\left( \frac{u}{R_{AdS}} \right)^2 f(u) d\eta^2
\label{unflavmetr}\\
F_{(6)}&=&Q_c \left( \frac{u}{R_{AdS}} \right)^4 dx_0 \wedge
dx_1\wedge dx_2 \wedge dx_3  \wedge du \wedge d\eta\\
e^\phi &= &\frac{2\sqrt2}{\sqrt3 Q_c}\,\,,
\quad\qquad R_{AdS}^2=\frac{15}{2}
\eea
with:
\be
f(u)=1-\left( \frac{u_\Lambda}{u} \right)^5
\ee
where $Q_c$ is proportional to the number of color D4 branes and, to avoid a conical singularity at the origin,  the coordinate $\eta$
needs to be periodic:
\be\label{per theta}
\eta \sim \eta + \delta \eta\,\,,\qquad\qquad
\delta\eta=\frac{4\pi R_{AdS}^2}{5 u_\Lambda}\,\,.\
\ee
We also define $M_\Lambda$ as the typical mass scale below which the
theory is effectively four-dimensional:
\be\label{MLAMBDA}
M_\Lambda=\frac{2\pi}{\delta\eta}=
\frac{5}{2} \frac{ \ u_\Lambda}{R_{AdS}^2}
\ee

Several interesting questions about the  gauge theory dual to this background
can be addressed by analyzing the closed string spectrum of  non-critical string theory on (\ref{unflavmetr})~\cite{KS2}. Despite the fact that one expects large stringy corrections to affect every calculation on the non-critical gravity side, the results obtained in \cite{KS2} are at least the same order of magnitude as those given by experiments or lattice calculations, showing therefore that also in this case the approximation to the complete string theory being considered is not meaningless.  This encouraged us to push  the study of this background a little  further, and introduce  fundamental degrees of freedom in the gauge theory. 

Since the gauge theory is non-supersymmetric, there seem to be two
 different ways to  do this, that is by adding additional D4 or 
 D5-branes, which we will generically call flavor probe  branes. 
 In fact, while both ways of proceeding  provide fundamental degrees 
 of freedom, the interpretation of the resulting gauge theory turns 
 out to be quite  different in the two cases. It seems natural to 
 argue that the low-energy limit of the gauge theory contains 
 massless fundamentals only in the case where these were 
 added through D4-branes. The reason is straightforward,
  and relies on the same argument that gives non-supersymmetric, 
  four-dimensional, pure  Yang-Mills as the low-energy limit of 
  the five-dimensional theory living on the wrapped D4 color 
  branes \cite{Wpure, KS1}. As mentioned above, to project out 
  the massless gaugino from the low-energy spectrum, we have to 
  impose anti-periodic boundary conditions over the thermal 
  cycle \cite{Wpure}. But when we add flavor D5-branes 
  (which obviously  need to wrap the $S^1$), the same mechanism
  gives mass to the  quarks  of the four-dimensional theory, leaving us with the massless degrees of freedom of a four-dimensional, non-supersymmetric pure Yang Mills theory plus massive quarks and gauginos.
  As a confirmation, we will see in section \ref{flavor D5} that indeed when a large number of D5-branes is added to the color D4-branes, the new background that is obtained  has the same form as the $AdS_6$ black hole (\ref{unflavmetr}), but with different coefficients, signaling that new massive UV states are being integrated out when going to the IR. Nonetheless, the addition of D5-branes is still interesting because it allows us to explore QCD  at  finite temperature.

For the addition of D4-branes, instead, the enforcing of anti-periodic boundary conditions over the circle should not affect the fundamental degrees of freedom which are introduced by the addition of the new flavor branes.  This is certainly the case in critical string theory, where the addition of flavor branes transverse to the $S^1$ circle provides massless quarks in the low-energy theory \cite{KMMW,SS}. We argue that the same is also true  for the present non-critical string theory. Therefore we expect that the introduction of flavor D4-branes provides an alternative scenario for the study of a string dual to QCD.

In both the D4 and D5 cases the flavor branes need to fill the whole four-dimensional  Minkowski space and stretch along the radius $u$ up to infinity. This last condition makes the gauge coupling on the flavor branes very small, freezing  its gauge group to a rigid symmetry group, that is the QCD $U(N_f)\times U(N_f)$ flavor group.

One relevant feature of the analysis of QCD in terms of a string dual is the ability to reproduce the spontaneous chiral symmetry breaking of the $U(N_f)\times U(N_f)$ flavor group into the diagonal  $U(N_f)$ when all the quarks are massless.
 
The two factors of the flavor group are obtained by introducing two different stacks of flavor branes, that is one of branes and the other one of anti-branes, in such a way that strings hanging between a color D4-brane and a flavor brane transform as quarks, while strings hanging between a color D4 and a flavor anti-brane transform as anti-quarks. The overall system of flavor branes is uncharged, as in \cite{KM,SS,BCCKP}. The chiral symmetry breaking is achieved by a reconnection of the brane-anti-brane pairs \cite{SS}.

\newsection{Probe D4 flavor branes  on the $AdS_6$ black hole}\label{probe D4}

As argued above, the most natural way of adding 
(few) flavors to the
$AdS_6$ black hole background \cite{KS1} is to include D4-probe branes
extended along the Minkowski directions and stretching to infinity in the
radial one. The picture is very similar to the 
approach to QCD in \cite{SS} where flavors are added to the ten-dimensional background of \cite{Wpure} by considering D8,
$\rm{\bar{D8}}$ flavor branes. The
main difference is the absence of the transverse $S^4$ of the critical case,
which plays no physical role.

We consider the  action:
\be
S_{D4}=-T_4 \int d^5x \,e^{-\phi} \sqrt{-\det(\hat g + 2\pi \alpha' F)} 
+ T_4 \, \tilde a \int {\cal P}(C_{(5)})
\label{generalaction}
\ee
where $\hat g$ stands for the pullback of the metric on the probe worldvolume.
It has to be pointed out that this action does not account for all the degrees
of freedom of the system. In particular, there are strings connecting the 
flavor D4 and $\bar{{
\rm D4}}$ branes, but we lack a reasonable description of an effective action for them.
Nevertheless, we do not expect them to destabilize the setup.
We leave the WZ coupling in the action as a direct
generalization of the well known one in the critical setup, although, as
far as we know, it is not well understood what should be written in this two-derivative approximation to the non-critical setup.
One should take $\tilde a =0$ if this WZ coupling is not present at
all whereas $\tilde a =1$ would be the direct naive generalization from the
ten-dimensional theory.

We consider  $x_i, \eta$ as the worldvolume coordinates of the 
probe brane and allow for $u=u (\eta)$. We find the lagrangian 
density (prime denotes derivative with respect to $\eta$):
\be
{\cal L}=-T_4 e^{-\phi} \left[
\left(\frac{u}{R_{AdS}}\right)^4
\sqrt{\left(\frac{u}{R_{AdS}}\right)^2 f(u) + \left(\frac{R_{AdS}}{u}\right)^2
f^{-1}(u) u'^2} - a \left(\frac{u}{R_{AdS}}\right)^5 \right]
\ee
where we have defined a new constant $a = \frac{2}{\sqrt5} \tilde a$
in order to simplify the notation.
Since the lagrangian does not explicitly depend on $\eta$,
the quantity $u'\frac{\partial{\cal L}}{\partial u'} - {\cal L}$
is a constant, and we can write:
\be
\left(\frac{u}{R_{AdS}}\right)^5 
\left[ \frac {f(u)}{\sqrt{f(u)+
\left(\frac{R_{AdS}}{u}\right)^4 \frac{u'^2}{f(u)}}}-a\right]=
\left(\frac{u_0}{R_{AdS}}\right)^5 \left[f^\frac12 (u_0) -a \right]
\ee
where we have defined $u_0$ as the minimal value of the radial 
coordinate reached by the probe brane. 
Therefore, the classical static 
embedding\footnote{$a<1$ is required for consistency.} is given by:
\be\label{embed}
\eta_{st}(u)=
\int_{u_0}^u \frac{(u_0^5 f^\frac12 (u_0)
-a u_0^5 +a u^5)\ du}{\left(\frac{u}{R_{AdS}}\right)^2
f(u) \sqrt{u^{10}f(u)-(u_0^5 f^\frac12(u_0)-a u_0^5 +a u^5)^2}}
\ee
Notice that tuning the constant of integration
$u_0$  modifies the boundary conditions at infinity, namely $\eta(\infty)$.
On general grounds, we may expect that different $u_0$ account for
different dual theories and we interpret this constant as being related
to the constituent mass  of the quarks, as will be discussed below in section \ref{sect: mass}.
In particular, for $a=0$, $\eta(\infty)|_{u_0=u_\Lambda}
=\delta \eta /4$ (where $\delta\eta$ is as defined in (\ref{per theta})) and 
$\eta(\infty)|_{u_0 \to \infty} \to 0$, as in \cite{SS}.
In the probe approximation which we use here, $u_0\neq u_\Lambda$ is a stable solution of the equation of motion
and as will be shown shortly the fluctuations around it are non-tachyonic. However, it might be that because the D4 flavor branes are codimension one, by going beyond
the probe approximation the backreaction with the $F_6$ field strength associated with the flavor branes may become important and a matter of concern.\footnote{We thank
Igor Klebanov for pointing this out to us.}

In the following, we  consider small fluctuations around this
embedding which correspond to the physical particles composed by
the fundamentals. The analysis is similar to that in \cite{SS}
but here we keep the constant of integration $u_0$ generic. In section
\ref{sect: mass} we will discuss its relation to the mass of the 
quarks. Indeed, we can define a mass parameter $m_q$ as the 
 energy of a
string connecting the color and flavor branes \cite{Kinar} (stretching from the 
horizon to $u_0$), {\it i.e.} 
\be\label{quarkmass}
m_q = E_s=\frac{1}{2\pi\alpha'}\int_{u_\Lambda}^{u_0} \sqrt{-g_{00}g_{uu}}du=\frac{1}{2\pi\alpha'}\int_{u_\Lambda}^{u_0} f^{-\frac12}(u)du .
\ee

\subsection{Fluctuations of the gauge field}
\label{sect: Ffluct}

The equations of motion for the gauge fields living  on the D4 flavor brane, that follow from the probe action (\ref{generalaction}), are:
\be
\partial_m\left(\sqrt{-\det \hat{g}}\,F^{mn}\right)=0 \, .
\ee
By splitting the field strength into its four-dimensional and internal components, we can write:
\begin{align}
&u^4\gamma^{\frac{1}{2}}\partial_\mu F^{\mu\nu}-\partial_u (u^4\gamma^\frac{1}{2}F^{\nu u})=0\label{eqngauge1}\\
\label{eqngauge2}&\partial_\mu F^{\mu u}=0\, ,
\end{align}
where we have defined
\be
\gamma = \frac{u^8}{f(u) u^{10} 
- (u_0^5 f^\frac12(u_0)-a u_0^5 +a u^5)^2}\, .
\ee
At the same time, we can expand the gauge field in the following way
\be\label{gaugedec}
\begin{split}
A_\mu (x^\mu, u)& = \sum_n B_\mu^{(n)} (x^\mu) \psi_{(n)} (u) \\
A_u(x^\mu,u)&= \sum_n \varphi^{(n)}(x^\mu) \phi_{(n)} (u)
\end{split}
\ee
and for convenience we also define
\be
F_{\mu\nu}^{(n)} (x^\mu) = \partial_\mu B_\nu^{(n)}-
\partial_\nu B_\mu^{(n)}
\ee
Under the decomposition (\ref{gaugedec}), equation (\ref{eqngauge2}) reads:
\be
\sum_n\left( \tilde{m}_n^2\varphi^{(n)}(x^\mu)\phi_{(n)}(u)-(\partial_\mu B_{(n)}^\mu(x^\mu))\partial_u\psi_{(n)}(u)   \right)=0
\ee
where $\tilde{m}_n$ is the four-dimensional mass of the field $\varphi^{(n)}$: $\eta^{\mu\nu}\partial_\mu \partial_\nu \varphi^{(n)}=\tilde{m}_n^2 \varphi^{(n)}$. We see immediately that choosing the Lorentz gauge $\partial_\mu B^\mu_{(n)}=0$  for the four-dimensional gauge fields sets to zero all the massive scalars $\varphi^{(n)}$. On the contrary,  the massless field $\varphi^{(0)}$ survives, regardless of the value of the free parameter $u_0$ appearing in $\gamma$. 

Let us consider now the other equation of motion (\ref{eqngauge1}). Substituting (\ref{gaugedec}) in, it is easy to show that (\ref{eqngauge1}) fixes the form of $\phi_{(0)}$ to be
\be
\phi_0\sim \frac{\gamma^\frac12}{u^2}\,\,,
\label{phi(0)}
\ee
and also imposes the defining equation for the modes $\psi_{(n)}(u)$:
\be
-\gamma^{-\frac12} \partial_u (u^2  \gamma^{-\frac12}
\partial_u \psi_{(n)}) =
R_{AdS}^4 m_n^2 \psi_{(n)}
\label{vectoreq}
\ee
where $m_n$ is the mass of the four-dimensional vector field $B^{(n)}_\mu$: $\eta^{\nu\rho}\partial_\nu \partial_\rho B^{(n)}_\mu={m}_n^2 B^{(n)}_\mu$. The last ingredient we still miss to make sense of the expansions (\ref{gaugedec}) are orthonormality conditions for the modes $\psi_{(n)}$ and $\phi_{(0)}$. The natural choice  is to require that the kinetic terms in the four-dimensional effective action are canonically normalized, and therefore the conditions we are after, read
\ba
C \,(2\pi\alpha')^2\int du \gamma^{\frac12} \psi_{(n)}\psi_{(m)} &=&\delta_{nm}\,\,,
\label{vectornorm1}\\
C  \,(2\pi\alpha')^2\int du  \frac{u^2}{R_{AdS}^4} \gamma^{-\frac12} 
\phi_{(0)} \phi_{(0)} &=&  1\,\,,
\label{vectornorm2}
\ea
where:
\be
C=T_4 e^{-\phi} R_{AdS}\, .
\ee
It is straightforward to show that the orthogonality condition (\ref{vectornorm1}) is consistent with equation (\ref{vectoreq}), and that the mode $\phi_0$ in (\ref{phi(0)}) is normalizable.

Near infinity, the set of orthonormal functions $\psi_{(n)}$ 
solving equation (\ref{vectoreq}) behaves as:
\be
\lim_{u\to \infty} \psi_{(n)} \sim C_1  + \frac{C_2}{u^2}
\ee
Normalizability (\ref{vectornorm1})
fixes $C_1=0$. Moreover, the worldvolume coordinate $u$ describes
separately the two halves of the brane and, in order to be smooth at
the origin, one has to require that $\psi$ is an odd or even function
on the worldvolume of the probe.
 These two conditions can be simultaneously satisfied only for certain
values of $m_n$, generating, then, a discrete spectrum. We leave its
analysis for section \ref{sect: numerics}.

Finally, the effective four-dimensional flat space action for the fluctuations of the gauge field $A^m$ living on the probe D4-brane can be written as:
\be\label{gaugeactionbis}
S=-\int d^4x \left[ \frac12 \partial_\mu\varphi^{(0)}
\partial^\mu\varphi^{(0)}+\sum_{n\geq 1} \left( \frac14
F_{\mu\nu}^{(n)}F^{\mu\nu\;(n)}+\frac12
m_n^2 B_\mu^{(n)}B^{\mu\;(n)}\right)\right]\,\,.
\ee
The field $\phi_{(0)}$ can be associated to  
$\partial_u \psi_{(0)}$ where $\psi_{(0)}$ is the zero mode of equation
(\ref{vectoreq}); the associated four-dimensional field $\varphi^{(0)}$ corresponds to the Goldstone boson.
As in \cite{SS}, one can also go to the $A_u=0$ gauge
where there are no $\phi_{(n)}$ and the Goldstone boson gets encoded
in the boundary ($u\to \infty$) value of $A_\mu$.
Notice that this massless mode exists for any value of $u_0$.
We will comment on this in section \ref{sect: mass}.

\subsection{Fluctuations of the embedding}

We now consider small fluctuations around the static embedding
(\ref{embed}), which are
scalars in the four-dimensional theory:
\be
\eta = \eta_{st} (u) + \xi (x^\mu,u)
\ee
Keeping up to quadratic terms in $\xi$ and discarding a total derivative, the
expansion of (\ref{generalaction}) yields: 
\be\label{5dactscal}
S_{D4}=-\frac12 C \int d^4x du 
\left[ \gamma^{-\frac32} \frac{u^4}{R_{AdS}^8} (\partial_u \xi)^2
+ \gamma^{-\frac12} \frac{u^2}{R_{AdS}^4} \eta^{\mu\nu}\partial_\mu \xi \partial_\nu \xi
\right]
\ee
The equation of motion for the scalar $\xi$ reads then:
\be\label{eqnscal}
R_{AdS}^4\gamma^{-\frac{1}{2}}u^2\eta^{\mu\nu}\partial_\mu\partial_\nu \xi+\partial_u(\gamma^{-\frac{3}{2}}u^4\partial_u\xi)=0
\ee
As for the vector field above, we make an expansion of the fluctuation field:
\be\label{scalardec}
\xi = \sum_{n=1}^{\infty} {\cal U}^{(n)}(x^\mu) \sigma_{(n)} (u) 
\ee
where the functions $\sigma_{(n)} (u)$ are defined by the second order equation 
\be
-u^{-2} \gamma^{\frac12} \partial_u ( u^4 \gamma^{-\frac32}
\partial_u \sigma_{(n)}) = R_{AdS}^4 m_n'^2 \sigma_{(n)}
\label{scalareq}
\ee
which follows directly from the equation of motion (\ref{eqnscal}). Here $m'_n$ are the possible mass eigenvalues for the four-dimensional scalars ${\cal U}^{(n)}$: $\eta^{\mu\nu}\partial_\mu \partial_\nu {\cal U}^{(n)}=m'^2_n {\cal U}^{(n)}$.

By substituting the expansion (\ref{scalardec}) into the action (\ref{5dactscal}), one finds the effective action for a tower of four-dimensional massive scalar mesons:
\be
S=\frac12 \int d^4x \sum_n\left[(\partial_\mu {\cal U}^{(n)})^2
+ m_n'^2 ({\cal U}^{(n)})^2\right]
\ee
where the normalization of the fields $ \sigma_{(n)}$ has been chosen in order for the kinetic term to be canonically normalized:
\be\label{normsigma}
C\int \gamma^{-\frac12} \frac{u^2}{R_{AdS}^4}  \sigma_{(m)}
\sigma_{(n)} = \delta_{mn} 
\ee

It is now easy to see that the asymptotic behavior of $ \sigma_{(n)}$ is of the form:
\be
\lim_{u\to \infty} \sigma_{(n)} \sim C_1  + \frac{C_2}{u^6}\,\,,
\ee
so normalizability (\ref{normsigma}) requires $C_1=0$ and, again, the mass spectrum $m'_n$ is discrete.

\subsection{Numerical analysis and discussion}
\label{sect: numerics}

In this section we proceed to a numerical study of equations (\ref{vectoreq})
and (\ref{scalareq}). 

Starting with the vectors (\ref{vectoreq}), notice that two kinds of boundary 
conditions can be imposed on the functions $\psi_{(n)}(u)$ at the bending 
point of the probe brane $u=u_0$, which lead to different parity and charge
conjugation for the corresponding mesons (we use $J^{CP}$ notation):
\ba
\partial_u \psi_{(n)}(u)\big|_{u=u_0}=0\quad &\rightarrow& \quad 1^{--} \qquad
\textrm{(odd $n$)}\\
\psi_{(n)}(u_0)=0\quad &\rightarrow& \quad 1^{++} \qquad
\textrm{(even $n$)}
\ea
The arguments to deduce $P$ and $C$ are completely analogous to those in
\cite{SS} and will not be repeated here. Notice, however, that we are assuming a
WZ coupling on the worldvolume of the flavor brane to the Hodge dual of the
RR-form, {\it i.e.} $\int_{D4} {}^*F_{(6)} \omega_5(A)$ where $d\omega_5 = F^3$.

\begin{figure}[tb] 
   \includegraphics[width=0.45\textwidth]{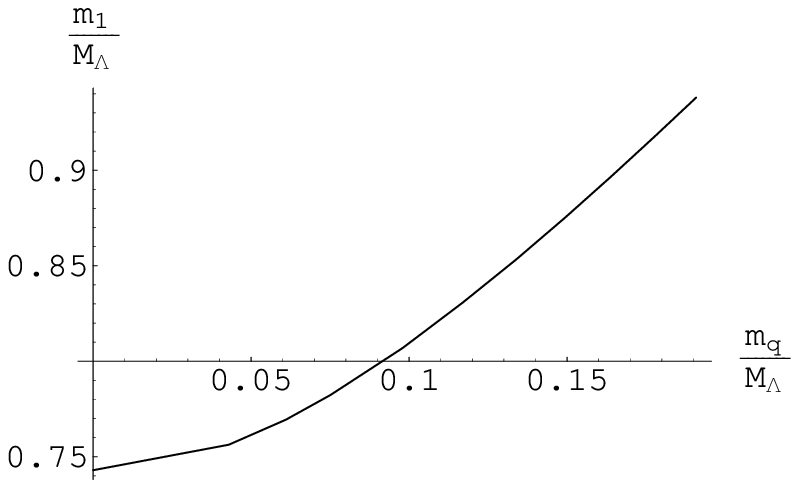} 
   \includegraphics[width=0.45\textwidth]{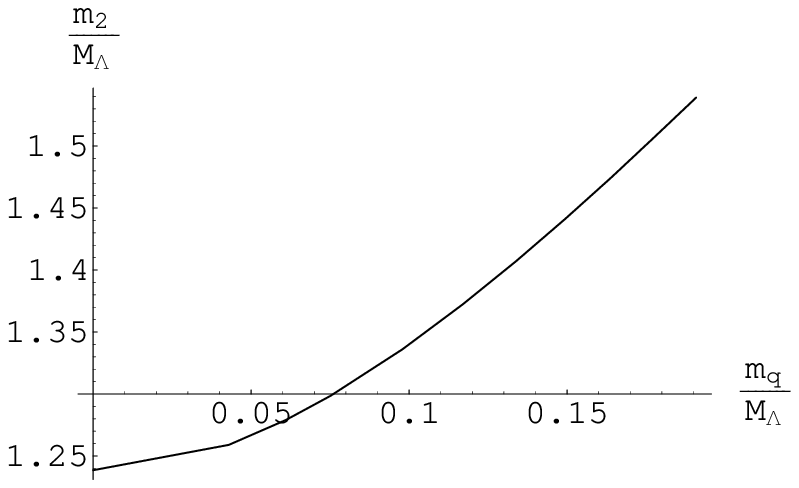} 
   \caption{Plots of the masses of the two lightest vector mesons
   varying the constant $u_0$. $a=0$ has been taken.}
   \label{fig: massplot}
\end{figure}

\begin{figure}[htb] 
   \includegraphics[width=0.45\textwidth]{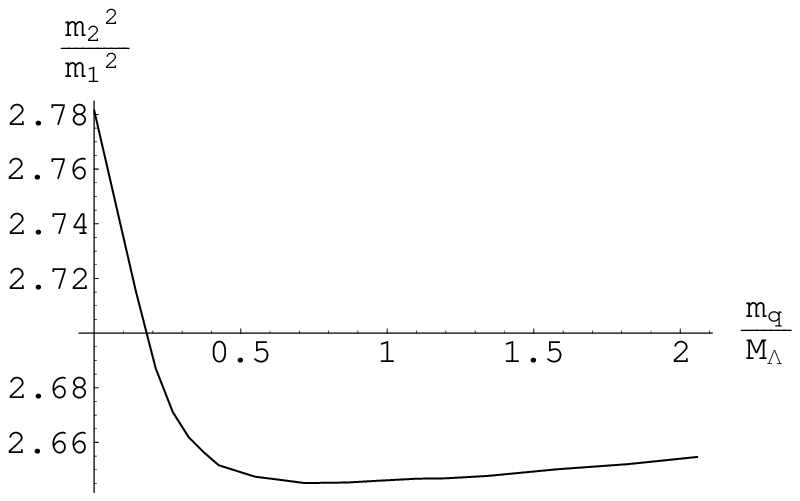} 
   \includegraphics[width=0.45\textwidth]{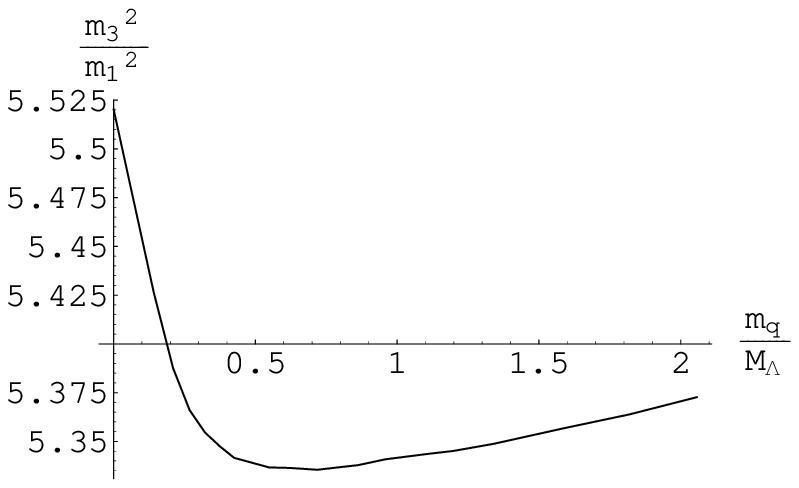} 
   \caption{Quotients of the squared masses of the vector mesons 
   for different
   values of the bare mass of the quarks
   (we have set $a=0$). The experimental values
   of such quotients for the corresponding QCD mesons are:
   $\frac{m_2^2}{m_1^2}=2.51$, $\frac{m_3^2}{m_1^2}=3.56$. }
   \label{fig: quotients}
\end{figure}

\begin{figure}[htb] 
   \includegraphics[width=0.45\textwidth]{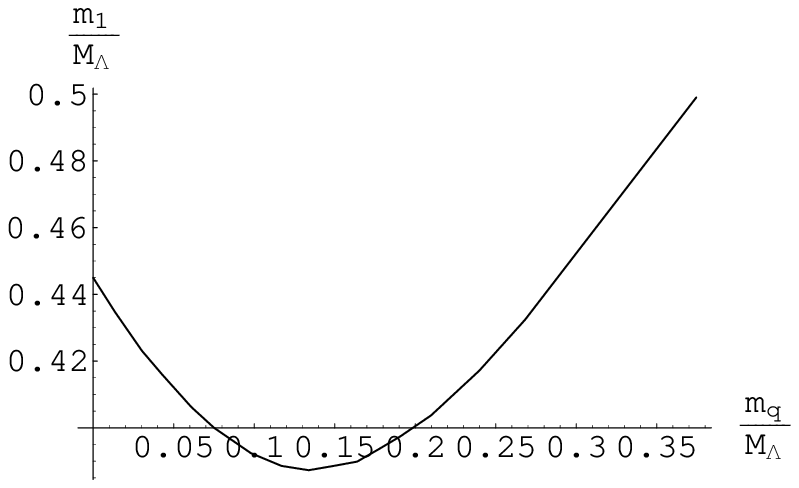} 
   \includegraphics[width=0.45\textwidth]{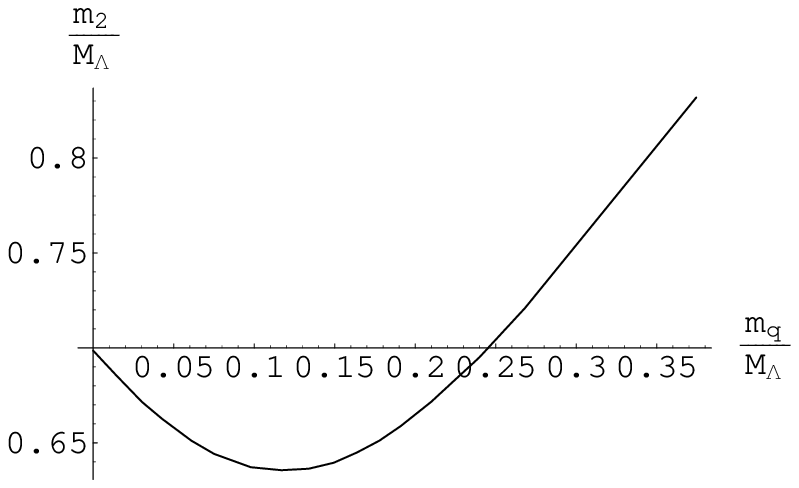} 
   \caption{Plots of the masses of the two lightest vector mesons computed
   with $\tilde a =1$. 
   }
   \label{massplotwitha}
\end{figure}

Let us start by setting $a=0$. Then, the results are summarized in figures
\ref{fig: massplot} and~\ref{fig: quotients}.
In particular, notice that the meson masses grow 
monotonically with the quark
mass parameter. On the contrary, see figure \ref{massplotwitha},
if we set $a\neq 0$, we find the  feature that the
meson masses decrease when $m_q$ grows (at least for small $m_q$).

\begin{figure}[thb] 
\begin{center}
   \includegraphics[width=0.45\textwidth]{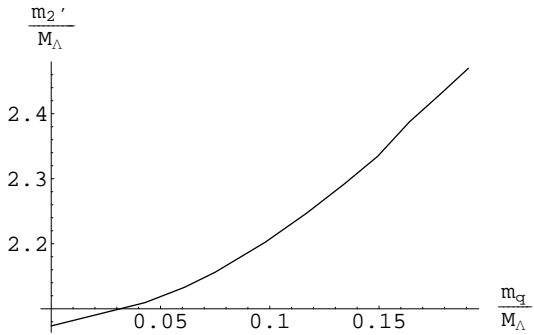} 
   \end{center}
   \caption{The mass of the next to lightest scalar meson vs.
   the quark mass ($a=0$).}
   \label{scmeson}
\end{figure}

The study of scalar and pseudoscalar mesons using equation (\ref{scalareq})
is a bit more subtle. The choice of $u$ as a worldvolume coordinate
and $\eta$ as the excited scalar is ill-defined near the tip of the probe brane,
where the boundary conditions have to be imposed. In fact, this formalism does
only allow us to find half of the modes: those in which the point at 
$u=u_0$ is fixed, {\it i.e.} the odd excitations on the brane, which correspond
to $0^{--}$ mesons, yielding the $m_n'^2$ for even $n$ 
(Figure \ref{scmeson} shows the computed $m_2'^2$ for different $m_q$
taking $a=0$). 
In particular, we
cannot compute the mass of the lightest one, which is of physical importance
since it would correspond to the $a_0(1450)$ meson \cite{SS}.
A similar technical problem was discussed in \cite{Nunez:2003cf}, where
it was overcome by an appropriate change of the choice of the coordinates
describing the worldvolume and the excited scalar. In this case, we have
not been able to find such a convenient change for general $u_0$ and the
formalism gets too complicated. 
On the other hand, for the case $u_0 = u_\Lambda$ (with $a=0$) it is indeed
possible to find a convenient change of coordinates, in analogy
with \cite{SS}. Details are reported in appendix \ref{app: det}.
The masses of the lightest states with $u_0=u_\Lambda$ and
$a=0$ are:
\be
\begin{split}
&m_1^2=0.55 M_\Lambda^2\qquad \, m_2^2=1.53 M_\Lambda^2\qquad \, m_3^2=3.04 
M_\Lambda^2\qquad\,  m_4^2=5.06 M_\Lambda^2\qquad \ldots \\
&m_1'^2=2.28 M_\Lambda^2\qquad m_2'^2=4.30 M_\Lambda^2\qquad 
m_3'^2=6.82 M_\Lambda^2\qquad m_4'^2=9.83 M_\Lambda^2\qquad \ldots
\end{split}
\label{scmasses}   
\ee

\paragraph{Discussion} \  \\

As all non-critical string models where the action has been truncated 
to terms with two derivatives, the model we are considering is 
expected to suffer from relevant curvature  corrections. Nonetheless, 
symmetry arguments suggest that the underlying $AdS$ structure of the 
space should provide some protection and make corrections smaller 
than expected, as we already noticed to be the case for the unflavored 
solution analyzed in \cite{KS2}. For this reason it is still 
interesting to compare our predictions for the spectrum of mesons
 with the experimental values, and with the values obtained in the 
 context of critical string theory in  \cite{SS}. We summarize the 
 results in table \ref{mass ratios}.
\begin{table}[h]
\begin{center}
\begin{tabular}{|c|c|c|c|}
\hline
 \ & {\rm experiment} & \textrm{D4-D8 model} & \textrm{non-critical D4-D4}
\\
\hline
$m^2_2 / m^2_1$ & 2.51 & 2.4 & 2.8 \\ \hline
$m^2_3 /m^2_1$ & 3.56 & 4.3 & 5.5 \\ \hline
$m'^2_1 / m^2_1$ & 3.61 & 4.9 & 4.1 \\ \hline
$m^2_2/m'^2_1$ & 0.70 & 0.49 & 0.67 \\ \hline
\end{tabular}
\end{center}
\caption{Comparison of some experimental meson mass ratios with the
results obtained from the 
model of \cite{SS} and the non-critical model
with $a=m_q=0$. \label{mass ratios}}
\end{table}

We have to notice, though, that curvature  corrections to the present model are 
not the only kind of corrections we expect to arise. 
 Just as the critical dimension  models in~\cite{KMMW,SS}, 
 the thermal modes, that is the KK modes surviving the anti-periodic 
 boundary conditions projection over the $S^1$ circle $\eta$,  have 
 masses of the same order of magnitude as the lightest glueballs of 
 the four-dimensional Yang-Mills theory. Therefore meson masses and 
 scattering amplitudes will get contributions from loops containing 
 these spurious modes, which do not belong to the spectrum of pure 
 Yang-Mills. For the models in \cite{KMMW,SS}, there are even more 
 KK modes which do not belong to the spectrum of pure Yang-Mills: 
 they come from the  internal space over which ten-dimensional string 
 theory is compactified. One of the advantages in using non-critical 
 strings (even though partially overshadowed by the effects of stringy
  corrections) is that these modes are not present simply because 
  there is no additional internal space over which the non-critical 
  theory needs to be compactified.

\subsection {Spinning open strings}
\label{sect: sos}

Apart from the spectrum of the lightest mesons of the gauge theory, we can 
use the string/gauge duality to study the spectrum of mesons with very large 
spin. Let us consider the dynamics of an open string that has its endpoints 
attached to a probe flavor brane, and that rotates on a plane inside 
four-dimensional Minkowski.  The interpretation of this string is that 
of a  meson in the gauge theory, composed by a quark-antiquark pair. 
By analyzing  the relationship between the energy and angular momentum 
of the spinning open string, we expect to be able to study the Regge 
trajectory of mesons in a QCD-like theory \cite{KMMW, KPSV,Paredes:2004is}.

We leave most of the details of the calculation for appendix \ref{app: spinning}, and collect here only the relevant results and comments. We consider, for clarity, a string centered at the origin of four-dimensional Minkowski space, spinning with constant angular velocity $\omega$ on the $x^1,x^2$ plane, which we parameterize as
\be
ds^2=dR^2+R^2 d\phi^2
\ee

The classical expressions for the energy and angular momentum of the open string are given in (\ref{E}) and (\ref{J})
\begin{align}
&E=T_q \int  
d\sigma (1+\rho^2)^\frac{2}{5}\sqrt{\frac{R'^2+\frac{4}{25}
\frac{R_{AdS}^4}{u_\Lambda^2} \frac{\rho'^2}{(1+\rho^2)^{7/5}}
 }{1-\omega^2R^2}} \label{Esosintext}
\\[7pt]
&J=T_q\,
\omega \int  d\sigma R^2(1+\rho^2)^\frac{2}{5}\sqrt{\frac{R'^2+
\frac{4}{25}\frac{R_{AdS}^4}{u_\Lambda^2} \frac{\rho'^2}
{(1+\rho^2)^{7/5}} }
{1-\omega^2R^2}} \label{Jsosintext}
\end{align}
where
\be\label{Tq}
T_q\equiv \frac{1}{2\pi\alppr}\left(\frac{u_\Lambda}{R_{AdS}}\right)^2 = \frac{3}{5\pi} M_\Lambda^2
\ee
is the effective tension for a string stretching close to the horizon of the $AdS_6$ black hole (\ref{unflavmetr}) \cite{KPSV}. In general, these expressions are largely affected by  stringy corrections coming from the vibration of the string around its classical configuration. Nonetheless there is a particular limit where a semiclassical approximation is reliable, that is when the angular momentum $J$ of the string is really large. In this case, the angular velocity of the string is really small and the distance between its two endpoints becomes very large. This configuration is really close to the static string representing the Wilson loop for two external quarks \cite{BISY2}, and therefore the string can be effectively described as three straight segments: two vertical ones connecting the probe D4-brane to the horizon of the black hole, and an horizontal one lying on the horizon of the black hole. The string behaves like a large string rotating on the horizon of the black hole, with two massive quarks attached at its endpoints.

In this limit it is easy to show that the energy and angular momentum of the string are given by (\ref{Esos}) and (\ref{Jsos}) respectively
\begin{align}\label{EJ}
&E=\frac{2T_q}{\omega}\left(  \frac{\sqrt{1-\omega^2 R_0^2}}{\omega R_0} +\arcsin  (\omega R_0) \right)\\
& J=\frac{T_q}{\omega^2}\left(  \omega R_0  \sqrt{1-\omega^2 R_0^2} +\arcsin  (\omega R_0) \right)
\end{align}
where the interquark distance $2R_0$ is determined in terms of the quarks dynamical  mass and angular velocity $\omega$ through (\ref{R0})
\be
R_0=\frac{m_q}{2T_q}\left( \sqrt{1+\frac{4T_q^2}{m_q^2 \omega^2}} -1 \right)
\ee

Therefore, in the limit of $J$ very large where the semiclassical approximation is valid,  the energy and angular momentum of the spinning open string  satisfy a Regge law
\be\label{EMJ}
E^2=2\pi T_q J =\frac{6}{5} M_\Lambda^2 J 
\ee

\begin{figure}[htb] 
   \centering
   \includegraphics[width=0.5\textwidth]{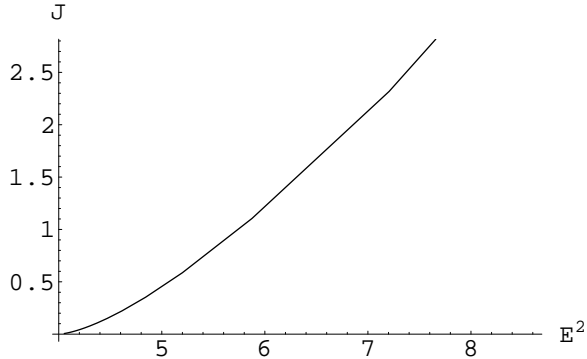} 
   \caption{The plot shows an extrapolation to the small angular momentum region of formulae (\ref{Esosintext}) and (\ref{Jsosintext}). Our approximate analysis shows that in this region the $(E^2,J)$ trajectory deviates from the linear Regge behavior. Moreover, it is evident that there is a non-zero negative intercept.}
   \label{E2/J}
\end{figure}

Even though our analysis is not strictly reliable away from the large $J$ region, it is still interesting to extrapolate the $(\mathrm{energy})^2$/angular momentum trajectory from (\ref{Esosintext}) and (\ref{Jsosintext}) to lower values of $J$. The result is shown in figure \ref{E2/J}. If the semiclassical analysis  is at least approximately valid outside the large $J$ region,  the arguments above predict a deviation from the linear Regge behavior.
 A qualitative analysis of the  small $J$ limit is also reported at the end of appendix \ref{app: spinning}.

Note that (\ref{EMJ}) means that the mass scale of the mesons on the Regge trajectories is the same as that of the 
massive mesons  discussed in section 3.3, namely, $M_\Lambda$. This is not the case in the critical model of \cite{SS}
where  $T_q\sim M_\Lambda^2 g^2_{YM}N_c$ which means that only for $g^2_{YM} N_c\sim 1$ the mass scales of the  mesons associated with the fluctuations of the probe branes and the spinning strings, can be the same. This obvioulsy implies
curvature of order one, the same as in the non-critical model.   

\subsection{On the mass of the quarks}
\label{sect: mass}
We want to end this section with a brief discussion  about the  mass of the quarks as extracted
from the analysis of the gravity background plus probe branes versus the notions of quarks masses from QCD and the quark model. Let  us first remind ourselves  the latter. In the QCD action  bare masses for the various flavored quarks are introduced, which
 upon renormalization turn into  physical masses, which sometimes are referred to as current algebra masses. For instance the $u$ and the $d$ quarks have masses of few $MeV$. Let us denote this mass as $m_q^{QCD}$.
 This mass parameter is very significant for many QCD processes. Here we focus only on one issue, that is the mass of the pion.
 It is well known  that the mass square of the pion is proportional to  $m_q^{QCD}$ as follows
\be\label{mpi}
m_\pi^2 \sim (m_u^{QCD} +  m_d^{QCD})
\ee
 where $m_u^{QCD}$  and $m_d^{QCD}$ are the physical masses of the $u$ and the $d$ quarks  respectively. 
  In addition, in the context of the quark model or potential models, one  defines a
  constituent mass parameter which is roughly for the $d$ and $u$ quarks one third
  of the mass of the proton, namely around $300$ $MeV$.

In the gravity plus probe brane model we consider, there are three pieces of information that can teach us which role the quark mass plays in this setup.

\begin{itemize}  
 
\item
 The  spectrum of  massive mesons\\
 We found that on top of the Goldstone bosons there are massive vector 
 and pseudoscalar mesons and that  their masses increase 
monotonically\footnote{For $a\neq 0$ this is the case only from some value of $m_q$ on.} with the mass 
parameter $m_q$, defined in (\ref{quarkmass}) as the energy of the string that stretches between  the lowest 
point of the probe brane and the wall.
In fact, from some mass scale of the order of $M_\Lambda$, 
the meson masses grow linearly in $m_q$ (see figure \ref{fig: linear}),
and the slope is bigger for higher states\footnote{
 In the supersymmetric case of \cite{KMMW}, the meson masses also grow linearly in the
 quark mass.}.
\begin{figure}[tb] 
   \includegraphics[width=0.45\textwidth]{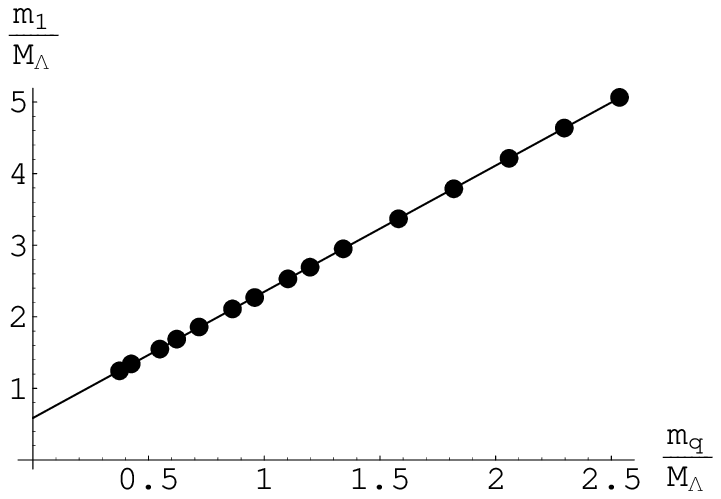} 
   \includegraphics[width=0.45\textwidth]{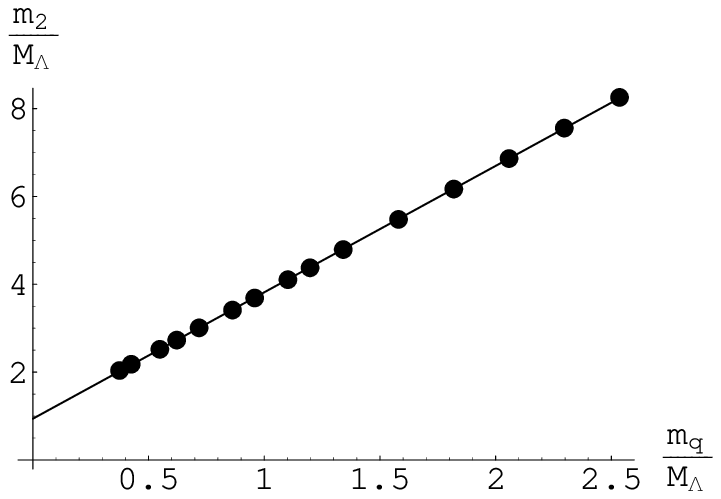} 
   \caption{Asymptotically, the meson masses grow linearly in the
   parameter $m_q$. Here we present the plot for the two lightest
   vector mesons.}
   \label{fig: linear}
\end{figure}
We therefore conclude that this mass parameter $m_q$ is 
related to the constituent mass of the quark.  
\item
The spectrum of large spin mesons\\
As we have seen above, the spinning string that describes a meson of large spin, has a U shape,   stretching from a probe brane vertically to the wall, along the wall and then again vertically to a probe brane (not necessarily the one it started from). 
The energy of  each of the  two vertical parts, which  is  exactly $m_q=E_s$,
 contributes to the total mass of the meson as a relativistic 
 spinning particle, as can be seen when we rewrite the energy $E$ of (\ref{EJ}) 
 in the following form
\be
E= \frac{2m_q}{ \sqrt{1-w^2 R_0^2}}+\frac{2T_q}{\omega}\left(  \arcsin  (\omega R_0) \right)
\ee
This result agrees with our interpretation of $m_q$ as related to the constituent mass of the quarks.

\item 
The spectrum of Goldstone bosons\\
 As mentioned in section \ref{sect: Ffluct}, the mass of the Goldstone bosons 
 is always zero in our model, even when $u_0\neq u_\Lambda$ and therefore $m_q$ as defined in (\ref{quarkmass}) is nonzero. 
 On the contrary, we know that if the quarks 
have a bare mass, the
chiral $U(1)$ is not a symmetry of the lagrangian and the Goldstone boson should
also acquire a mass, proportional to 
$\sqrt{m_q^{QCD}}$  for small $m_q^{QCD}$.
This behavior was first found in a holographic setup in 
\cite{Babington:2003vm}. 
 Thus  we see that our gravity 
 model seems to describe quarks with vanishing
current algebra masses $m_u^{QCD} =  m_d^{QCD} =0$.
To the best of our knowledge creating a current algebra mass 
 in the models based on a brane anti-brane configuration is still an open question.
 Note that a non-vanishing constituent mass is possible
even if the current algebra mass is zero.
However
  since in QCD one cannot vary the constituent mass without changing the
current algebra mass, there is an obvious puzzle in these results.
In~\cite{SS}, following arguments of
\cite{Sugimoto:2004mh}, it was argued that in the D4-D8-$\rm{\bar{D8}}$ critical
setup a bare mass for the quarks should be related to an expectation value
for the tachyon field coming from the open strings joining the flavor branes and
anti-branes. Since the effective action we wrote does not account for these
modes, it cannot describe this feature satisfactorily. 
It is reasonable to conjecture that in the full stringy description of this system, 
there would not be a massless mode when $u_0\neq u_\Lambda$, but unfortunately
we lack a better understanding of this point.

\end{itemize}


\newsection{Backreaction with large\,-$N_f$ D4 flavor branes}\label{flavor D4}

It would be very interesting to be able to go beyond the probe 
approximation $N_f\ll N_c$. This would allow us to evaluate also those 
quantities of QCD which strongly depend on the dynamics of fundamental 
 quarks. In this section we take a first step in this direction,  
 by studying what happens when many flavor D4-branes are added to 
 the $AdS_6$ black hole background of \cite{KS1, KS2}.

The action we start from reads (we  take $\alpha'=1$)
\be
S=\frac{1}{2k_6^2}\int d^6x \sqrt{-g} \left\{ e^{-2\phi} 
\left(R+4\, \partial_\mu \phi\, \partial^\mu \phi +4 \right) 
-\frac{1}{2 \cdot 6!} \,F_{(6)}^2  \right\}+ S_{\mathrm{flavors}}
\label{backraction}
\ee
where $k_6$ is the 6-dimensional Newton's constant, and the action 
for the flavor D4-branes is given by:
\be
S_\mathrm{flavors}=T_4\sum^{N_f}\left(-\int_{\mathcal{M}_5} d^5x\, 
e^{-\phi}\sqrt{-\hat{g}_5}+\tilde a \int_{\mathcal{M}_5} 
\mathcal{P}(C_{(5)})\right)
\label{Sflavore}
\ee
Here $\hat{g}_5$ is the induced metric over the D4-brane worldvolume 
$\mathcal{M}_5$, while $\mathcal{P}(C_{(5)})$ is the pull-back of the RR 
5-form potential over the D4-branes worldvolume. $\tilde a$ is a constant 
which fixes the relative strength of the DBI and CS terms
as in eq. (\ref{generalaction}). 
Unlike the probe case here due to the particular configuration we consider, with the flavor branes smeared over the $\eta$ compact direction, 
 it is clear that one has to  
set $\tilde{a}$  to zero.
Notice also that taking a non-trivial value for $\tilde a$ would 
generate terms in the action which are not periodic under shifts
 $\delta\eta$ of $\eta$ (cf.~(\ref{per theta})), which is definitely 
 not acceptable.

We look for backgrounds of the form
\be\label{ansatz metric}
ds^2=e^{2\lambda} dx^2_{1,3}+dr^2+e^{2\tilde{\lambda}}d\eta^2
\ee
and take the most symmetric assumption, that is all functions 
$\lambda$, $\tilde{\lambda}$ and the dilaton $\phi$ can only depend 
on  the radius $r$. The color D4-branes wrap the circle $\eta$, while
 the flavor D4-branes are points on this circle. In order to simplify 
 the following calculations, we  take a $U(1)$ symmetric distribution 
 of the flavor D4-branes, that is we  smear them uniformly  over 
 the circle, in a way analogous to \cite{BCCKP}. 
 The DBI term, reads then
\be
-T_4\sum^{N_f}\int d^5x\,e^{-\phi}\,\sqrt{-\hat{g}_5}=
-T_4\int d\eta \frac{N_f}{\delta\eta}\int d^5x\,e^{-\phi}\,
\sqrt{-\hat{g}_5}=-\frac{T_4\,N_f}{\delta\eta}\int d^6x\,
e^{-\phi-\tilde{\lambda}}\,\sqrt{-g_6}
\ee

It is now trivial to write the equation of motion for the 6-form field strength
\be
d*F_{(6)}=0
\ee
which  therefore requires
\be\label{eqn F_{(6)}}
*F_{(6)}=Q_c
\ee
 where $Q_c$ is a constant. It is straightforward to show that  
 the following  5-form potential is a good solution to (\ref{eqn F_{(6)}})
\be
C_{(5)}= \beta(r) dx^0\wedge dx^1\wedge dx^2\wedge dx^3\wedge  d\eta 
\ee
with 
\be
\beta'(r)=Q_c\,e^{4\lambda+\tilde{\lambda}}\ee

We are now ready to write the effective action for the ansatz (\ref{ansatz metric}). We skip a few intermediate steps (integration by parts and some algebra), and display the result
\be
S_{eff}=\frac{4\pi \mathrm{Vol}(\mathbb{R}^{1,3})}{k_6^2}\int dr \, e^{4\lambda+\tilde{\lambda}-2\phi}\left(3\lambda'^2+2\lambda'\tilde{\lambda}'-4\lambda'\phi'-\tilde{\lambda}'\phi'+\phi'^2+1-\frac{Q_f}{4}e^{\phi-\tilde{\lambda}}-\frac{Q_c^2}{8}e^{2\phi}\right)
\ee
where we defined $Q_f=2\frac{T_4 k_6^2}{\delta\eta}N_f$.

As in \cite{KS1}, we define a deformed dilaton 
\be
\varphi=-2\phi+4\lambda+\tilde{\lambda}
\ee
and a new radial coordinate $\rho$ such that $d\rho= -e^{-\varphi}dr$. The effective lagrangian reads, then, (a dot stands for derivation with respect to $\rho$)
\be
\mathcal{L}=\dot{\varphi}^2-4\dot{\lambda}^2-\dot{\tilde{\lambda}}^2+4 e^{2\varphi}-\frac{Q_c^2}{2}e^{\varphi+4\lambda+\tilde{\lambda}}-Q_f \,e^{\frac{3}{2}\varphi+2\lambda-\frac{1}{2}\tilde{\lambda}}
\ee
The  zero-energy condition reads
\be\label{0nrg D4}
\dot{\varphi}^2-4\dot{\lambda}^2-\dot{\tilde{\lambda}}^2=4 e^{2\varphi}-\frac{Q_c^2}{2}e^{\varphi+4\lambda+\tilde{\lambda}}-Q_f \,e^{\frac{3}{2}\varphi+2\lambda-\frac{1}{2}\tilde{\lambda}}
\ee
while after going back to the original dilaton $\phi$, the equations of motion read
\begin{align}
&\ddot{\phi}=-2e^{-4\phi+8\lambda+2\tilde{\lambda}}+\frac{3}{4} Q_c^2e^{-2\phi+8\lambda+2\tilde{\lambda}}+\frac{3}{4} Q_fe^{-3\phi+8\lambda+\tilde{\lambda}}\label{eom phi}\\
&\ddot{\lambda}=\frac{Q_c^2}{4} e^{-2\phi+8\lambda+2\tilde{\lambda}}+\frac{Q_f}{4} e^{-3\phi+8\lambda+\tilde{\lambda}}\label{eom lambda}\\
&\ddot{\tilde{\lambda}}=\frac{Q_c^2}{4} e^{-2\phi+8\lambda+2\tilde{\lambda}}-\frac{Q_f}{4} e^{-3\phi+8\lambda+\tilde{\lambda}}\label{eom lambda tilde}
\end{align}

These equations are really involved and finding an explicit solution has 
proven a useless effort, up to now. But there is a special class of 
solutions which can be easily derived, those with a constant dilaton 
 $\phi=\phi_0$. The equation for $\phi$ (\ref{eom phi}) turns then into 
 an algebraic constraint
\be\label{alg 1}
\begin{aligned}
&e^{2\phi_0}=\frac{8}{3 Q_c^2} & \qquad &Q_f=0\\
&e^{\tilde{\lambda}}=\frac{3Q_f e^{-\phi_0}}{8 e^{-2\phi_0}-3 Q_c^2}&\qquad &Q_f\neq 0
\end{aligned}
\ee 
The case $Q_f=0$ was considered in \cite{KS1,KS2}, and leads to
the $AdS_6$ black hole solution (\ref{unflavmetr}). 
On the other hand, if $Q_f \neq 0$, one finds an $AdS_5 \times
S^1$ solution which can be obtained by performing T-duality
along the $S^1$
to the solution found in \cite{KM}:
\be
e^{\phi_0}=\frac{2}{\sqrt{3} Q_c}\,\,,\qquad
R_{AdS}^2=6\,\,,\qquad
R_{S^1}=\frac{\sqrt3}{2}\frac{Q_f}{Q_c}\,\,.
\ee


\newsection{Flavored $AdS_{n+1}\times S^k$ backgrounds and black holes}
\label{sect: general}

The same line of reasoning as the one we followed in the preceding section 
can be generalized to a much wider class of non-critical string backgrounds. 
It has been found in~\cite{KS1}, that the two-derivative truncation of 
the non-critical string theory action admits solutions of the form 
$AdS_{n+1}\times S^k$. They can be interpreted as the backgrounds 
generated by stacks of color D$(n-1)$-branes in  $D=n+k+1$ dimensions. 
Here we consider the effect of including also space-filling D$(n+k)$ 
flavor branes. This analysis was made in \cite{alish} for the 
particular case $D=6$.
We will not worry about the stability of the solutions 
we find:  we think it is convenient to keep a general formalism, even 
though some of the solutions considered could prove to be unstable
due to open string modes.

We start from the two-derivative approximation to the $D=n+k+1$-dimensional non-critical string action in the string frame (with $\alppr=1$)
\be\label{S general}
\begin{split}
S=&\frac{1}{2 k_D^2}\int d^D x\sqrt{-g}\left(e^{-2\phi}(R+4\partial_\mu \phi \partial^\mu \phi+ 10-D)-\frac{1}{2(n+1)! }F_{n+1}^2\right)+\\&-N_f T_{D-1}\int d^Dx e^{-\phi}\sqrt{-g}
\end{split}
\ee
where the $e^{-2\phi}(10-D)$ term is the usual central charge appearing when working off criticality, and the last term is the contribution from the $N_f$ uncharged flavor branes. We look for configurations that have the following structure:
\be\label{ansatz general}
ds^2=e^{2\lambda}dx^2_{1,n-2}+e^{2\tilde{\lambda}}d\eta^2+dr^2+e^{2\nu}d\Omega_k
\ee
where the $D(n-1)$-branes extend along the $x^0,\ldots,x^{n-2}$ directions, are wrapped around the compact direction $\eta$, and are transverse to a $k$-sphere. We take all functions, including the dilaton $\phi$,  to depend only on $r$. The equation of motion for the $(n+1)$-form $F_{n+1}$ reads $d*F_{n+1}=0$, which is solved by
\be\label{F general}
F_{n+1}=Q_c e^{(n-1)\lambda+\tilde{\lambda}-k\nu} dr\wedge dx^0 \wedge dx^1\wedge \ldots\wedge dx^{n-2}\wedge d\eta
\ee
After substituting our ansatz (\ref{ansatz general}) and (\ref{F general}) into (\ref{S general}), we find the effective action
\be
\begin{split}
S\sim \int d\rho&
\left(\dot{\varphi}^2-(n-1)\dot{\lambda}^2-\dot{\tilde{\lambda}}^2-
k\dot{\nu}^2+k(k-1)e^{2\varphi-2\nu}+(10-D)e^{2\varphi}+\right.\\&
-\left.\frac{Q_c^2}{2}e^{\varphi+(n-1)\lambda+\tilde{\lambda}-k\nu}-
Q_fe^{\frac{3}{2}\varphi+\frac{n-1}{2}\lambda+\frac{1}{2}\tilde{\lambda}
+\frac{k}{2}\nu}\right)
\end{split}
\ee
where we have defined $Q_f=2k_D^2 T_{D-1} N_f$,
introduced a generalized dilaton:
\be
\varphi=-2\phi +(n-1)\lambda+\tilde{\lambda}+k\nu
\ee
and used a
 new radial coordinate such that
\be
d\rho=-e^{-\varphi}dr
\ee
The equations of motion coming from this effective action read
\begin{align}
\begin{split}
&\ddot{\phi}=-\frac{10-D}{2}e^{-4\phi+2(n-1)\lambda+
2\tilde{\lambda}+2k\nu}+\frac{n-k+1}{8}Q_c^2e^{-2\phi+2(n-1)
\lambda+2\tilde{\lambda}}+\\&\qquad+\frac{D+2}{8}Q_f 
e^{-3\phi+2(n-1)\lambda+2\tilde{\lambda}+2k\nu}
 \end{split}\label{1 gen}\\
&\ddot{\lambda}=\frac{Q_c^2}{4}e^{-2\phi+2(n-1)\lambda+2\tilde{\lambda}}+\frac{Q_f}{4}e^{-3\phi+2(n-1)\lambda+2\tilde{\lambda}+2k\nu}\label{2 gen}\\
&\ddot{\tilde{\lambda}}=\frac{Q_c^2}{4}e^{-2\phi+2(n-1)\lambda+2\tilde{\lambda}}+\frac{Q_f}{4}e^{-3\phi+2(n-1)\lambda+2\tilde{\lambda}+2k\nu}\label{3 gen}\\
&\ddot{\nu}=(k-1)e^{-4\phi+2(n-1)\lambda+2\tilde{\lambda}+2(k-1)\nu}-\frac{Q_c^2}{4}e^{-2\phi+2(n-1)\lambda+2\tilde{\lambda}}+\frac{Q_f}{4}e^{-3\phi+2(n-1)\lambda+2\tilde{\lambda}+2k\nu}\label{4 gen}
\end{align}

Finding the general solution to this coupled system of differential 
equations is far too difficult to hope to be able to get a sensible 
answer. Nonetheless, we can explicitly derive a class of special 
solutions, that is those with a constant dilaton $\phi=\phi_0$. 
This makes the first equation (\ref{1 gen}) into an algebraic 
equation relating the value of $\phi_0$ with the function~$\nu$. 
In particular when $k\neq n+1$, (\ref{1 gen}) requires also $\nu$ to 
be constant $\nu=\nu_0$.  For $k=n+1$ one could in general have a 
non-constant radius for the $k$-sphere. Nonetheless, we will stick 
to the simpler case $\nu=\nu_0$ also when $k=n+1$. Again, this makes 
the fourth equation (\ref{4 gen}) into another constraint on $\phi_0$ 
and $\nu_0$. 
We now denote the radius of the $S^k$ and the string coupling as follows
\be
R_S\equiv e^{\nu_0} \qquad g_s \equiv e^{\phi_0}
\ee
After some algebra, one finds that $R_S^2$ satisfies an equation of degree $k+2$
\be\label{nu0 gen}
\begin{split}
&(10-D)R_S^{2(k+2)}+(D+2)(k-1)R_S^{2(k+1)}+\\&-\frac{2}{n+2}\frac{Q_c^2}{Q_f^2}\left((10-D)R_S^{2}-(n-k+1)(k-1)\right)^2=0
\end{split}
\ee
and the string coupling  is given by
\be\label{phi0 gen}
g_s =\frac{2}{Q_f}\frac{10-D-(n-k+1)(k-1)R_S^{-2}}{n+2}
\ee
Notice that when $Q_c\sim Q_f$, all the coefficients in equation (\ref{nu0 gen}) are order one, and therefore the radius of the $k$-sphere is a finite number while, from equation (\ref{phi0 gen}), the string coupling is order $1/Q_f$. 

It is now straightforward to solve equations (\ref{2 gen}) and (\ref{3 gen}) for $\lambda$ and $\tilde{\lambda}$
\begin{align}
&\lambda=-\frac{1}{n}\log \left(\frac{1}{b_1}\sinh \left(b_1\sqrt{hn}\rho\right)\right)+a_1\rho\\
&\tilde{\lambda}=-\frac{1}{n}\log \left(\frac{1}{b_1}\sinh \left(b_1\sqrt{hn}\rho\right)\right)-(n-1)a_1\rho
\end{align}
where $h$ is a parameter we have introduced to simplify formulas
\be
h=\frac{Q_c^2}{4}e^{-2\phi_0}+\frac{Q_f}{4}e^{-3\phi_0+2k\nu_0}\ ,
\ee
whereas $a_1$ and $b_1$ are two constants of integration. The
other two 
have been fixed by taking a suitable origin for $\rho$ and rescaling 
the coordinates $x^0,x^1,\ldots, x^{n-2}$ and $\eta$.  
A relation holds between the two remaining constants 
$a_1$ and $b_1$, and it comes from imposing the zero-energy condition
\bea
&&\dot{\varphi}^2-(n-1)\dot{\lambda}^2-\dot{\tilde{\lambda}}^2-
k\dot{\nu}^2= \\
&&=k(k-1)e^{2\varphi-2\nu}+(10-D)e^{2\varphi}
-\frac{Q_c^2}{2}e^{\varphi+(n-1)\lambda+\tilde{\lambda}-k\nu}-
Q_fe^{\frac{3}{2}\varphi+\frac{n-1}{2}\lambda+\frac{1}{2}\tilde{\lambda}
+\frac{k}{2}\nu}\nonumber
\eea
The result is the following relation
\be
a_1=b_1\sqrt{\frac{h}{n}}
\ee

We found, therefore, a one-parameter family of solutions which are 
characterized by the dilaton and radius of the sphere transverse to 
the D$(n-1)$-branes being constant. As we will readily show these 
solutions are black hole deformations of  $AdS_{n+1}\times S^k$ 
backgrounds which are obtained by taking $b_1=0$. Let us start from 
this simpler case. If we set $b_1=0$, the expressions for $\lambda$ 
and $\tilde{\lambda}$ read:
\be
\lambda=\tilde{\lambda}=-\frac{1}{n}\log\left(\sqrt{hn}\,\rho\right)=\sqrt{\frac{h}{n}}e^{2\phi_0-k\nu_0}\,r
\ee
where an integration constant has been fixed in order to cancel 
a constant term. Thus for $b_1=0$ the solution is an 
$AdS_{n+1}\times S^k$ space, similar to those obtained in \cite{KS1}. 
The radius of the $AdS$ part is given by
\be
\frac{n(n+2)}{ R_{AdS}^2}=10-D +\frac{k^2-1}{R_S^2}
\ee
where $R_S^2$ is as given in (\ref{nu0 gen}). 

Let us now consider several classes of special cases:
\begin{itemize}
\item
The case that there is no  sphere at all  in the transverse directions, namely, where  the metric is of the form
$ ds^2=e^{2\lambda}dx^2_{1,n-2}+e^{2\tilde{\lambda}}d\eta^2+dr^2$ requires a special treatment.
Equations (\ref{1 gen})-(\ref{3 gen}) are still intact with  $k=0$, but 
obviously  (\ref{4 gen}) does not exist anymore.
The value of the string coupling is now
\be
g_s=\frac{D+2}{2D}\frac{Q_f}{Q_c^2}
\left [ \sqrt{1+ \frac{16D( 10-D)}{(D+2)^2}
(\frac{Q_c}{Q_f})^2}-1\right ]
\label{dilk=1}
\ee
and the radius of Anti de Sitter is given by
\be
R_{AdS}^2= \frac{4n}{g_s^2}[Q_c^2 + g_s^{-1}Q_f ]^{-1}
\label{radsk=1}
\ee

\item

For the case of critical superstrings, $D=10$ we get    for $k>1$  
\be
R_S^2= \left(   \frac{n+2}{24(k-1)}(g_s Q_c)^2\right )^{\frac {1}{k-1}}
\ee
and $R_{AdS}^2$ is related to $R_S^2$ in the following way
\be
\frac{n(n+2)}{R_{AdS}^2}- \frac {k^2-1}{R_{S}^2} = 0
\ee
and $g_s$ is given by
\be
g_s^\frac{k+1}{k-1} Q_f =2 \frac {(k-1)(k-n-1)}{n+2}\left( \frac{24}{Q_c^2}\frac {k-1}{n+2}\right)^{\frac {1}{k-1}}
\ee
In particular for the case with no flavors we get that the only solution is for 
$n=k-1=4$ which is the $AdS_5\times S^5$ solution. With flavors there is a solution only provided that $k>(n+1)$.
This implies that there is no  flavored $AdS_5\times S^5$ solution associated with space filling branes. On the other hand there are
solutions of the form 
 $AdS_4\times S^6$, $AdS_3\times S^7$ and  $AdS_2\times S^8$ which may be interpreted as the flavored near horizon limit of  
D2, D1 and D0 branes respectively.  
\item
A special class of  non-critical solutions is for the case of an $S^1$, namely,    $k=1$ where
\be
g_s = \frac{2(10-D)}{n+2 }\frac{1}{Q_f}\qquad 
R_{AdS}^2= \frac{D(D-2)}{10-D}   
\qquad 
R^2_S= 2\frac{10-D}{n+2}\frac{Q_c^2}{Q_f^2}
\ee
In particular for $k=1, n=4, D=6$ we get the KM solution \cite{KM}.

\item
For the special class  where $k = n+1$ we get
\be
g_s = \frac{2(10-D)}{n+2}\frac {1}{Q_f}= \frac {4(10-D)}{D+2}\frac {1}{Q_f}
\ee
The relation between the radii is
\be
\frac{1}{R_{AdS}^2}- \frac {1}{R_{S}^2} = \frac{10-D}{n(n+2)}
\ee

\end{itemize}
Let us now proceed and describe the corresponding black hole solutions where $b_1\neq 0$. We introduce a new radial coordinate~$u$
\be
\frac{u}{R_{AdS}}=e^\lambda
\ee
In terms of this new variable, the solution reads
\be
ds^2=\left(\frac{u}{R_{AdS}}\right)^2\left(dx^2_{1,n-2}+\left(1-\left(\frac{u_\Lambda}{u}\right)^n\right) d\eta^2\right)+\left(\frac{R_{AdS}}{u}\right)^2\frac{du^2}{1-\left(\frac{u_\Lambda}{u}\right)^n}+e^{2\nu_0}d\Omega_k
\ee
where $u_\Lambda$ is the position of the horizon of the black hole
\be
u_\Lambda^n=2 b_1 R_{AdS}^n
\ee
Notice that when $b_1>0$, the metric component $g_{\eta\eta}$ vanishes at $u=u_\Lambda$. Therefore the background is defined only for $u\geq u_\Lambda$ and to avoid a conical singularity at the horizon, we have to impose that the coordinate $\eta$ is periodic
\be
\eta\sim \eta+\frac{4\pi R_{AdS}^2}{n\,u_\Lambda}
\ee
Before closing let us mention three particular
 solutions that are relevant via holography to interesting 
 gauge theories.

\subsection{Near extremal flavored $AdS_6$}
\label{flavor D5}

This solution corresponds to adding uncharged spacetime filling branes
to the background of sections \ref{sect: AdS6} and \ref{probe D4}.
As argued in section \ref{sect: AdS6}, adding such D5-branes corresponds to 
adding fundamental flavors to a gauge theory at a finite temperature. 

\begin{figure}[htb] 
\begin{center}
   \includegraphics[width=0.45\textwidth]{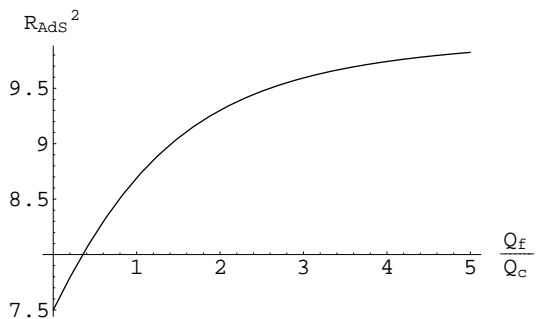} 
   \end{center}
   \caption{Variation of $R_{AdS}^2$ with the ratio of flavors to
   colors in the flavored $AdS_6$ black hole.}
   \label{fig: rads}
\end{figure}

One can read from (\ref{dilk=1}) and (\ref{radsk=1}) the values of the
dilaton and $R_{AdS}$:
\be
e^{\phi_0}=\frac{2}{3}\frac{Q_f}{Q_c^2}
\left(\sqrt{1+6\frac{Q_c^2}{Q_f^2}}-1\right)\,,\qquad
R_{AdS}^2=\frac{90}{12+\frac{Q_f^2}{Q_c^2}-
\frac{Q_f^2}{Q_c^2}\sqrt{1+6\frac{Q_c^2}{Q_f^2}}}\,\,.
\label{ads6values}
\ee
It is interesting that $R_{AdS}$ depends on the 
ratio of number of colors and flavors (see figure \ref{fig: rads}). 
In the context of the $AdS_5$ background, it was conjectured in reference
\cite{BCCKP} that certain values of $R_{AdS}$
may bound the stable gravity solutions, leading to a conformal window in the
field theory bounded by certain $N_f/N_c$ ratios.

\subsection{ Near extremal KM model}

There is a particularly interesting case, among all those considered 
in the main part of this section: the $AdS_5\times S^1$ black hole. 
The setup is  the same as the one in \cite{KM}: we consider a 
system of D3 and uncharged D5-branes in six dimensional non-critical 
string theory. But here, we compactify one of the gauge theory flat 
directions on a thermal cycle, in order to break supersymmetry. The 
resulting background is dual to a four-dimensional field theory at  
a finite temperature with fundamental flavors, which becomes effectively three-dimensional for high values of the temperature. This is the adaptation 
of the idea of Witten \cite{Wpure} to a theory in three dimensions with 
the important novelties that here  we consider the set-up in non-critical 
string theory, as it has been done in \cite{KS1,KS2}, and moreover that 
our starting  theory  contains fundamental degrees of freedom.

The background reads
\be\label{kmbh}
ds^2=\left(\frac{u}{R_{AdS}}\right)^2\left( dx^2_{1,2}+
\left(1-\left(\frac{u_\Lambda}{u}\right)^4\right)d\eta^2\right) 
+\left(\frac{R_{AdS}}{u}\right)^2\frac{du^2}{1-\left(\frac{u_\Lambda}{u}\right)^4}+R_{S^1}^2 d\theta^2
\ee
with $R_{AdS}=\sqrt{6}$, $R_{S^1}^2=\frac{4 Q_c^2}{3 Q_f^2}$, $u_\Lambda^4=2 b_1 R_{AdS}^4$ and $e^{\phi_0}=\frac{4}{3Q_f}$. The RR 5-form field strength reads
\be
F_5=Q_c\left( \frac{u}{R_{AdS}}  \right)^3 dx^0\wedge dx^1 \wedge dx^2 \wedge d\eta \wedge du
\ee
and its dual is $\partial \chi=Q_c\, d\theta$. The period of the compact direction $\eta$ is given by
\be
\eta\sim \eta+\frac{\pi R_{AdS}^2}{u_\Lambda}
\ee

\subsection{ Near extremal flavored $AdS_5$ model}
\label{Ads5}
The solutions above also include a non-extremal  
flavored $AdS_5$ model, generalizing the five-dimensional 
construction of \cite{BCCKP}. 
 The metric is the same as that in (\ref{kmbh}) apart 
from the fact that now the $S^1$ term  is missing.
The value of the string coupling 
is\footnote{Notice that the difference with the expressions in
\cite{BCCKP} comes from different definitions of $Q_c$ and $Q_f$.}
\be
e^{\phi_0}=\frac{7}{10}\frac{Q_f}{Q_c^2}\left [ 
\sqrt{1+ \left(\frac{20 Q_c}{7Q_f}\right)^2}-1\right ]
\ee
and $R_{AdS}^2$ is given by:
\be
R_{AdS}^2= \frac{\frac{400}{7} }
{\frac{100}{7}+\frac{Q_f^2}{Q_c^2}
[1-\sqrt{1+\left(\frac{20 Q_c}{7Q_f}\right)^2}]}
\ee


\newsection{ Gauge dynamics from the flavored near extremal  models }
\label{sect: phenom}

We can now study the phenomenology of four-dimensional gauge dynamics
from three models that were derived in the previous sections:
\begin{enumerate}
\item
The flavored near extremal $AdS_6$ model
\item
The near extremal KM model
\item
The flavored near extremal $AdS_5$ model.
 \end{enumerate}
The first two models are six-dimensional and  the last one is five-dimensional.
The construction of the first and the third  models can be viewed as either
following the path
\begin{displaymath}
AdS\quad \rightarrow \quad \textrm{near extremal $AdS$}\quad  \rightarrow \quad \textrm{flavored  near  extremal  $AdS$}
\end{displaymath}
or 
\begin{displaymath}
AdS\quad  \rightarrow \quad \textrm{flavored  $AdS$}\quad \rightarrow \quad   \textrm{near  extremal 
flavored  $AdS$}
\end{displaymath}
whereas the second model, since it does not have an unflavored 
predecessor,  necessarily follows
\begin{displaymath}
\textrm{flavored $AdS_5\times S^1$} \quad \rightarrow \quad   \textrm{near  extremal flavored $AdS_5\times S^1$}
\end{displaymath} 

Before extracting  gauge dynamical properties, let us identify the properties of the backgrounds that affect them.
The backgrounds are  characterized by 
  the $AdS$ radius, the value of the constant dilaton, the radius of the thermal circle,
 the flux of the ``colored'' RR form $\sim Q_c$ and the number of flavor branes $Q_f$. 
Table \ref{table: comparison} includes the values of these features of the models. For completeness we added also the values of the
unflavored cases.

\begin{table}[h]
\begin{center}
\begin{tabular}{|c|c|c|}
\hline
 \ & $g_s Q_c$ & $R^2_{AdS}$ 
\\ \hline\hline
unflavored $AdS_6$ & $\sqrt{\frac{8}{3}}$& $\frac{15}{2}$ \\ \hline\hline
flavored $AdS_6$ & $\frac{2}{3}\frac{Q_f}{Q_c}\left(\sqrt{1+6\frac{Q_c^2}{Q_f^2}}-1\right)$ &$ \frac{90 }{12+\frac{Q_f^2}{Q_c^2}
[1-\sqrt{1+6\frac{Q_c^2}{Q_f^2}}]}$ \\ \hline\hline
KM model & $\frac{2}{3}\frac{Q_c}{Q_f}$ & $6$ \\ \hline\hline
unflavored $AdS_5$ & $2$ &  $4$ \\ \hline\hline
flavored $AdS_5$ & $\frac{7}{10}\frac{Q_f}{Q_c}\left 
[ \sqrt{1+ (\frac{20 Q_c}{7Q_f})^2}-1\right ]$ &$ 
\frac{\frac{400}{7} }
{\frac{100}{7}+\frac{Q_f^2}{Q_c^2}
[1-\sqrt{1+\left(\frac{20 Q_c}{7Q_f}\right)^2}]}$\\ \hline\hline
\end{tabular}
\end{center}
\caption{Comparison of the 't Hooft parameter $g_sQ_c$ and the $AdS$ radius between the 
various models.\label{table: comparison}}
\end{table}

Note that as for any other non-critical model, here too the curvature is order one in units of $\alpha'$ and cannot be reduced
by taking either $Q_c$ or $Q_f$ to be large. Thus, as expected, we are facing 
for these models the  generic curvature problem. On the other hand we can always be in  a regime where the string coupling $g_s$ is small. Obviously for the first and third models this happens  by taking $Q_c$ to be large and for the second
$Q_f$ to be large. Note however that the claim is that this model \cite{KM} is valid only for $Q_c\sim Q_f$ and hence again taking large number of colors guarantees small string coupling.

At this point the first question we need to address is what phase  of the  gauge theory dynamics can be associated with each of these gravity models.
\begin{itemize}
\item
The starting point of the first model can be taken to be the unflavored conformal $AdS_6$ background. The latter is supposed to be
the dual of a fixed point of a non-supersymmetric five-dimensional gauge theory without fundamental quarks. The vanishing of the beta function may be due
to  the existence of adjoint matter or it might be the ``fixed point'' of the theory at the far UV where the theory is free. Since the latter has a set of infinitely many high spin conserved charges, it does not seem to correspond to the $AdS_6$ background and hence it can be only the former option. 
Note however that in five dimensions such fixed points are not known and may even be forbidden.
When we add flavors the existence of the flavored $AdS_6$  background means that there is still an  IR fixed point 
which presumably  will  move with respect to the unflavored one. 
 Next we turn on near extremality and dualize in fact  a   four-dimensional low energy 
effective gauge theory by using 
 the mechanism of \cite{Wpure} of taking the radius of the thermal cycle to be small.
The resulting gauge theory has a massless sector  which 
 is identical to that of the pure four-dimensional Yang Mills theory regardless of what nature the original fixed point had.
This sector will be realized in the IR theory in the form of glueballs.
In addition there are quarks with a mass of the order of the temperature which will form mesons in the IR theory.
Those of course are absent in the unflavored theory.  
 
\item
The second model is the KM model which before turning on near extremality is conjectured to describe
the IR fixed point of the $N=1$ SQCD. 
Moreover, there is no unflavored $AdS_5\times S^1$ solution and hence it is clear that unlike the first case here a vanishing
beta function can be achieved only with flavors.
If indeed this is the case, namely it  is the holographic dual of the  IR fixed point of the $N=1$ SQCD,
 then the near extremal solution corresponds to 
turning on finite temperature in the four-dimensional theory which  breaks both
supersymmetry and conformal invariance.
The limit of large temperature
is dual in the IR to a three-dimensional gauge theory.
 \item
The extremal case of the third model corresponds to a 
flavored four-dimensional non-supersymmetric gauge theory at a fixed point.
Based on the arguments given for the first case, we conjecture that it 
 corresponds to an IR  fixed point of QCD in the flavored case.  In four dimensions there is no problem with such a fixed point.
The black hole solution in this case corresponds to 
finite temperature four-dimensional QCD. Again in the limit of large temperature  we end up with a three-dimensional 
gauge theory with massless gauge fields and very  massive quarks.    
\end{itemize}

The next question that we would like to investigate   is what  the 
properties of the dual gauge theory which can be extracted from the string side of the gauge/string duality are.
 We will address here the issues of the 
Wilson line, the low lying glueball  and  meson spectra, which are extracted from the gravitational background, and the higher spin glueball and meson states which follow from closed and open spinning string configurations.

\subsection{The Wilson line}
The Wilson line, namely, a string that connects 
 an external
quark anti-quark pair in the background of a near extremal $AdS$ solution, was investigated in \cite{BISY1}
and \cite{BISY2}. The former paper discussed the case of a theory at finite
temperature while in  the latter 
the focus is on the low-energy gauge theory at one less dimension which, in particular, 
  was shown to admit
an area law behavior for the Wilson loop. One of the examples considered is three-dimensional pure Yang Mills associated with the $AdS_5\times S^5$ black hole. Actually, the  result applies to other near extremal 
backgrounds not necessarily of $AdS$ type like the near extremal D4 branes  background that admits a confining Wilson loop in four-dimensional YM theory.  
 Let us start with this case and then we will address the finite temperature behavior.
The same situation as in  \cite{BISY2} happens with the flavored near extremal solutions that we discuss here in the first and second models.
The string  
stretches from its  endpoints down to the wall and then along the wall and up again. 
The corresponding energy of such a string is given by \cite{KS2}
\begin{equation}
E= T_q L - 2 \kappa
+ \mathcal{O} \left( ( \log L)^\gamma e^{-\hat a L} \right)
\end{equation}
where $T_q$ is as defined in (\ref{Tq})
\be
T_q=  \frac{g_{00}(0)} {2\pi \alpha^\prime}=  \frac{1} {2\pi \alpha^\prime}
 \left( \frac{u_\Lambda}{R_{AdS}} \right)^2,
\ee
and for the $AdS_6$ case
$\hat a= \sqrt{5} \frac{u_\Lambda}{R_{AdS}^2}$, $\gamma$ is a positive constant
and  $\kappa \approx 0.309 \frac{u_\Lambda}{2 \pi} $. Note also that $u_\Lambda\sim M_\Lambda$ see (\ref{MLAMBDA}).

However, the situation here is very different from that of the unflavored case.
To see this imagine that we start with a small separation between the external quark and anti-quark and then gradually increase it.
Up to a distance $L_{break}$ we have a worldsheet connecting the worldlines of the two external quarks. At that point it is energetically favorable for the system to create a dynamical quark anti-quark pair, namely a string of vanishing length,  and break the worldsheet into two parts one connecting the original external quark
and the dynamical anti-quark and the second one connecting the dynamical quark to the external anti-quark. Now increasing further
the separation distance  between the original pair will not change this picture. 
The breaking length is determined from 
\be
T_{q} L_{break}\sim M_\Lambda \ \ \ \rightarrow \qquad L_{break}\sim \frac{1}{M_\Lambda}
\ee
where we made use of the fact that the energy of the ``dynamical meson'' $\sim M_\Lambda$ and that $T_{q}\sim M_\Lambda^2$.
This is obviously the stringy manifestation  of the 
 well known fact that the Wilson loop of a theory with fundamental quarks breaks up.

The Wilson loop of the third model for the case of small near extremality corresponds to that of four-dimensional QCD theory at finite temperature.
In \cite{BISY1} it was found that for distances of  order $L<<\frac{1}{{\cal T}}$, where ${\cal T}$ is the 
temperature,   the energy as a function of the distance and temperature
is given by
\be
E= \frac{R_{AdS}^2}{\Gamma(\frac{1}{4})} \frac{1}{L} [ 1+ c({\cal T}L)^4]
\ee
where $c$ is a positive constant independent of the parameters of the background,
whereas for $L>>\frac{1}{{\cal T}}$ the quarks become free  and the binding
energy vanishes. The latter case is due to the  screening caused by the 
thermal bath. The difference between our case and that of the near extremal $AdS_5\times S^5$ model  discussed in \cite{BISY1}, is just the value
of $R_{AdS}$ but the basic behavior is the same.

\subsection { The glueball spectrum}
The glueball spectra is extracted \cite{oz} from the spectrum of the 
fluctuations of the dilaton, the metric and the RR one-form which are 
computed in a linearized limit of the gravity action. 
In the framework of the unflavored non-critical $AdS_6$ black hole, this 
calculation was performed in \cite{KS2}.
Since we are interested in the four-dimensional case, we discuss here only the 
flavored near extremal  $AdS_6$ solution among the set of solutions we have found.
In fact, what is left over to do is just to introduce the  
modifications to the  results of  \cite{KS2}  due to the presence of 
the flavor term in the action. The equations of motion of the 
fluctuating fields are given in (3.2)-(3.4) of \cite{KS2}. The flavor 
term does not affect the equation of  the RR form 
but does modify that of the metric and the source term of 
the Laplacian of the fluctuation of the dilaton. Let's discuss the latter.
 The shift of the dilaton equation of motion takes the following form
\be
\nabla \delta \phi = 4 e^{\phi_0} \delta \phi\rightarrow 
\nabla \delta \phi =4(   e^{\phi_0} - 
Q_f e^{2\phi_0} )\delta\phi
\ee
where $\delta \phi$ is the dilaton fluctuation 
and  $e^{\phi_0}$ on the left-hand  and right-hand sides are  those of the unflavored and flavored solutions  respectively. 

Notice that there is a critical value of $Q_f$, or more precisely 
$\frac{Q_f}{Q_c}$, where the source term vanishes.  It is obvious 
from the equation above that the condition for that is that
$1-Q_f e^{\phi_0}=0$ which by inserting the value of the dilaton in
(\ref{ads6values}) 
yields $\frac{Q_f}{Q_c}=\frac{\sqrt3}{2}$.
For $Q_f$ larger than the critical value the sign of the source is flipped. The study of this case is left for future investigation. Here we
assume that $Q_f$ is below the critical value.
The corresponding spectra of the $n^{++}=0^{++}, 1^{++}$ and  are given by
 \begin{equation}   \label{eq:WKBphi}
M^2_{\mathbf{n^{++}}, \phi} \approx  \frac{A}{\beta^2} k (k +B_n) + O(k^0)
\quad
\textrm{with}
\quad 
\beta = \frac{4 \pi R_{AdS}^2}{5 u_\Lambda}.
\end{equation}
where $A$ is a numerical factor that depends on $\frac{Q_f}{Q_c}$ (for $Q_f=0 $ it is 39.66), and $B_0=5.02$,
$B_1= \frac{1}{2}$  as extracted from the dilaton, one-form and  metric respectively.
For  the derivation of the spectrum and a comparison with spectra deduced from 
both critical near extremal D4 branes as well as lattice computations see \cite{KS2}.  

\subsection{Spinning closed strings}
The spectrum of glueballs of spin greater than 2  cannot be determined 
from the gravity background but rather only from  spinning closed string
 configurations. 
The analysis of classical spinning closed strings in  confining 
backgrounds was done in the context of the critical backgrounds in 
\cite{PSV,Pons:2004dk,Leo} and for the non-critical 
unflavored $AdS_6$ in \cite{KS2}.
The situation here is very similar to the later. It is straightforward to 
show that the configuration similar to that of the form  (\ref{spineq})
(but fixing $\rho=0$)
is a solution of the classical equations of motion. 
In fact it was shown in \cite{KS2} that the conditions 
to have a spinning classical configuration are in one to 
one correspondence with the conditions for the background to have 
an area law Wilson loop.
This classical spinning string admits a Regge behavior with

\begin{equation}
J =\frac{ E^2}{4\pi T_q}
\ee

Next we can incorporate quantum fluctuations around the classical 
spinning  string configuration associated with the Regge trajectory. This in general leads to the famous intercept.  This incorporation  
was previously done for the critical confining backgrounds in \cite{PSV} and for
the non-critical $AdS_6$ black hole in \cite{KS2}. The result found there takes the form 
\begin{equation}
J =\frac{ (E-z_0)^2}{4\pi T_q} - \frac{3}{24} \pi + \Delta_{\textrm{f}},
\end{equation}
where $z_0$ is proportional to $u_\Lambda$ and $\Delta_{\textrm{f}}$
is the contribution of the massless and massive fermionic modes which we do not know how to determine.
From this expression we read the  bosonic intercept 
$\alpha_0= z_0^2/(4\pi T_q) - \frac{3}{24} \pi$.
Note that 
 in addition to the intercept there is also a term linear in $E$.
A difference between the result in the critical model 
\cite{Leo} and the non-critical one is in the proportionally
factor between $z_0$ and $u_\Lambda$.

\subsection{The meson spectrum} 
The spectrum of the pseudoscalar and vector mesons  was analyzed in  
detail in section~\ref{probe D4} using the probe D4 flavor 
branes. To derive the corresponding spectrum for the fully backreacted 
backgrounds of the flavored $AdS$ black hole solutions and the near 
extremal KM model  we can  follow a very 
similar approach.
Recall that the flavor term (\ref{Sflavore}) is
in fact the flavor brane worldvolume action. One should compute the
system of linear equations coming from the coupled oscillations of
the closed and open string degrees of freedom of an action of the
kind (\ref{backraction}). Nevertheless, up to  quadratic 
order, the $F_{\mu\nu}$ oscillations
 do not mix with those of the background, so
after integrating over the thermal circle we will 
find again an
 action  which 
yields a discrete spectrum of massive vector mesons similar to the 
one found in  section \ref{probe D4}. However, having
many flavor branes on top of each other, the $U(N_f)$ non-abelian action
has to be used.
Unlike what has been  found for the D4 probe case, 
if flavor is introduced with D5s, we do not expect to have 
Goldstone bosons since as it was mentioned above the fundamental 
quarks acquire a large mass and decouple. 
Indeed, one can check explicitly
that the mode analogous to the $\phi_{(0)}$ of section \ref{sect: Ffluct}
(see eq. (\ref{phi(0)})) is not regular at the origin for the case
of a flavor D5 and, therefore, it
is not in the spectrum.
In section \ref{probe D4} we have analyzed the dependence of the 
masses of the mesons as a function of the mass parameter $m_q=E_s$ that is 
associated with the location of the flavor brane. It is not clear 
how to introduce the analogous parameter for the fully backreacted 
backgrounds.
Obviously we  can also extract  the spectrum  of  the pseudoscalar mesons
from  the fluctuations of the embedding in a similar manner to 
what was done with the flavored probe branes.

To draw the holographic picture of the mesons of spin higher than one, 
one needs to study spinning string configurations
rather than the gravity modes. This type of  analysis was done 
in section \ref{sect: sos}  for the case of D4 probe branes  
 and resulted in a picture 
of spinning strings in flat spacetime with massive endpoints, which 
is very similar to what was derived earlier in \cite{KPSV}.
In our case we have space-filling branes so that the endpoints of 
the open strings can reside at any point in spacetime. This is 
obviously different from the case of the probes. As already  mentioned, 
it is also not clear how to introduce the mass parameter that was 
related to the
constituent mass in the probe scenario.




\newsection{Summary and open questions}
\label{summary}
In this paper we have addressed the issue of  the 
non-critical holographic duality of gauge theories that incorporate quarks in the fundamental representation. 
This was done  both by using flavor probe branes as well as by using certain fully backreacted gravity backgrounds.
For the probe analysis  we followed \cite{SS}. We put a set of $N_f$ D4 anti-D4 branes in the background of the near extremal
non-critical $AdS_6$ solution. We solved for the classical configuration of
these probe branes and studied the fluctuations
of both  the gauge fields on the brane as well as of the embedding coordinates. The spectra of these fluctuations translate 
into the spectra of the vector  and (pseudo) scalar mesons respectively. We compared the results to those measured in experiments as well as to the values found in \cite{SS} using a critical string model, and found that they are in reasonably good agreement with both.
An important issue that we address is the Goldstone bosons associated with the flavor chiral symmetry breaking.
We computed the dependence of the meson masses on the ``mass parameter'' which is the minimal  distance between the
probe brane and the ``wall'' of the background. 

We then moved on to  the second approach. We derived a class of flavored near extremal $AdS_{n+1}\times S^k$ solutions.
In particular, we wrote down the flavored $AdS_6$ and $AdS_5$ black hole solutions and the near extremal version of the KM model.
We addressed the question of what  gauge theory phases the gravity models describe.
We presented conjectures about this question which we backed up with the calculations of Wilson loops and 
the spectrum of  glueballs. We also discussed the meson spectrum.

Our perspective has been dual throughout the whole paper. On one side, 
we focused on the holographic description of dynamical flavors. 
It is worth pointing out advantages and disadvantages of the probe approach
presented in section \ref{probe D4} with respect to that of the so-called
$AdS$/QCD duality \cite{mesonlist}: in both cases a simplifying 
assumption is taken, which does not hold a priori.
 In our case it is the 
two-derivative approximation to non-critical string theory on a 
highly curved background. In \cite{mesonlist},
the flavored degrees of freedom are dualized by considering different fields
in an $AdS$ space imposed by hand (not coming from some set of Einstein 
equations) with an {\it ad hoc} IR cutoff in order to enforce confinement.
Having $AdS$, this approach allows a better holographic description of,
for instance, the quark masses, the quark condensate and the GMOR
relation. The understanding of these features is still an open problem for our model
and for the critical one of \cite{SS}. On the other hand, we
believe
 that the qualitative advantage  of our 
approach is the fact that we are building our (unjustified) 
approximation on a justified and first-principle theory: non-critical 
string theory, and one may hope that progress in understanding this theory will
naturally lead to an improvement of our approximation.
In this sense, our setup is similar to \cite{SS}. With respect to that work, the
main advantage of the non-critical construction
is the disappearance of KK-modes, which is replaced by the
problem of large curvature. Notice that trying to decouple KK-modes in
critical string theory also leads naturally to highly curved space-times.
It is appealing that all these different approaches lead to qualitative
agreement with several QCD features.
 This aspect relates to  the second 
perspective we took in the paper, that is to test how reliable the 
two-derivative approximation we employ is. 
We found, for example, that already 
in our approximation, we could identify the same mechanism as in 
\cite{SS} to provide a natural description of the spontaneous 
breaking of chiral symmetry for massless quarks. The different models also
lead to a rough matching of some experimental quantities. We want
to stress that these facts are pointing out that the idea of a
higher-dimensional description of QCD has a solid basis.

This project is just one step in the long journey of discovering the string theory dual of QCD.
There are still several conceptual  problems with the present approach as well as certain  open questions. Among these questions  we find:
\begin{itemize}   
\item
As we have mentioned several times throughout the paper,  the low-energy
gravity (or supergravity) limit of non-critical string theory 
is problematic since it is characterized by order one curvature which means that
higher order curvature corrections are 
a priori not negligible. In \cite{KS1} certain arguments were made to back up the conjecture that models
 of the form $AdS_{n+1}\times S^k$ are robust against higher curvature
 corrections, even though the corresponding radii are corrected. Here since we are discussing near extremal solutions  these arguments do not hold anymore. In fact it is known that already in the 
critical models the near extremal backgrounds get corrections.
Nonetheless, we believe that the property of the black hole solutions we found, of having a wall with a non-trivial string tension along it, which guarantees a confinement behavior
\cite{Kinar}, will not change due to higher order  curvature corrections. Proving this conjecture is left  to future investigation. 

\item
The identification of a mechanism that generates a current algebra mass to the quarks and hence renders the Goldstone bosons massive
is far from being understood.  
In particular we showed that the Goldstone bosons remain massless even when  a non-trivial quark mass parameter is introduced.
We intend to continue the investigation of  this question.
\item 
In addition to the mass of the quarks there are several properties on the gauge
theory side that have not been yet extracted from the gravity side with the
flavor branes. In particular,  the expectation value of $\bar\psi \psi$
 was derived in the 
model based on D6 branes \cite{KMMW2} together with its dependence on the quark mass. This has not been done yet for
models based on branes anti-branes configurations, such as those we have been considering.

\item
In recent years there has been an intensive effort to reveal the relation between
gauge theories at finite temperature and their holographic gravity duals, with particular interest in 
 the confining-deconfining phase transition etc.
So far everything was done in the context of critical gravity backgrounds. The derivation of a similar description 
for the non-critical scenarios  is an interesting question that we intend to address in the future.

\item
On top of the spectrum, one would also like to compute the decay rates of the various 
types of mesons. First steps toward this goal were
taken in \cite {SS2} where decay rates of low lying mesons were computed. 
In \cite{PSZ} an analysis of the decay processes of 
mesons is addressed using  the picture of spinning open strings. 
These two methods of computing decay rates can be
also applied to the non-critical models discussed in this paper.
\item
As we have mentioned above one can extract the mesonic spectrum from the 
fully backreacted solution in a similar manner to what we
did for the probe setups. We intend to come back to this issue and 
compute the corresponding meson spectra.

\end{itemize}

\acknowledgments

We would like to thank  Ofer Aharony for many 
fruitful discussions and his valuable comments on the manuscript. We thank Francesco Bigazzi, Aldo Cotrone and Elias Kiritsis for many useful discussions, and 
Carlos N\'u\~nez for valuable comments
on the manuscript. We thank Igor Klebanov for his comment and Martin Kruczenski for a discussion. 
J.S.~would like to thank Elias Kiritsis and the Ecole Polytechnique for
 their hospitality.
A great part of this work was done  while J.S has been a 
visitor of the Ecole Polytechnique.
The work of J.S was supported in part by the German-Israeli Foundation for
Scientific Research and by the Israel Science Foundation.
The work of  R.C. and A.P.  was partially supported by
INTAS grant, 03-51-6346, CNRS PICS \# 2530,
RTN contracts MRTN-CT-2004-005104 and
MRTN-CT-2004-503369 and by a European Union Excellence Grant,
MEXT-CT-2003-509661.

\appendix
\setcounter{equation}{0}
\renewcommand{\theequation}{\Alph{section}.\arabic{equation}}

\newsection{Scalar mesons with $u_0=u_\Lambda$}\label{app: det}

In this appendix we find the equation that allows to compute
the spectrum of scalar mesons in the non-critical D4-D4-$\rm{\bar{D4}}$ 
system in
the case when the flavor D4-$\rm{\bar{D4}}$ probes are extended along u
at a fixed value of $\eta$, {\it i.e} in the case when one takes
$u_0=u_\Lambda$ and $a=0$. The discussion closely follows that 
in \cite{SS}.

Let us define new coordinates for the $(u,\eta)$ plane
of the metric (\ref{unflavmetr}):
\be\label{newcoords}
u^5 = u_\Lambda^5 + u_\Lambda^3 r^2\,\,,
\qquad\qquad
\tilde \eta =\frac{2\pi}{\delta \eta}\eta=
\frac{5 u_\Lambda}{2R_{AdS}^2} \eta\,\,,
\ee
so that the period of $\tilde \eta$ is $2\pi$. The metric in the
$(u,\eta)$ plane reads, in the new coordinates:
\ba
\left( \frac{R_{AdS}}{u} \right)^2 \frac{du^2}{f(u)} +
\left( \frac{u}{R_{AdS}} \right)^2 f(u) d\eta^2=
\frac{4}{25}\,\frac{R_{AdS}^2 u_\Lambda}{u^3}
\left(\frac{u_\Lambda^2}{u^2}dr^2 + r^2 d\tilde \eta ^2
\right)= \nonumber \\
= \frac{4}{25}\,\frac{R_{AdS}^2 u_\Lambda}{u^3}
\left[(1-h\,z^2)dz^2+(1-h\,y^2)dy^2
-2h\,z\,y\,dz\,dy \right]
\ea
where we have defined:
\be
y=r\cos\tilde\eta\,\,,\qquad\qquad
z=r\sin\tilde\eta
\ee
and
\be
h = \frac {1}{r^2}\left( 1- \frac{u_\Lambda^2}{u^2}\right)
\ee

The probe brane is located at a fixed value of
the angle $\tilde\eta$, say,
$\tilde \eta=\frac{\pi}{2}$ so, in the new
coordinates, the brane is extended in $z$ at $y=0$. We want to 
study small fluctuations around this embedding so we now consider
$y(x^\mu,z)$ and just keep terms in $y$ up to quadratic order. It is
easy to verify that the action (\ref{generalaction}) is:
\be
S_{D4}=-\tilde T \int d^4x dz \left[ u^\frac32+
\frac{2}{25}\,\frac{R_{AdS}^4 u_\Lambda}{u^\frac72}\eta^{\mu\nu}\partial_{\mu} y\,\partial_{\nu} y
+\frac12\,\frac{u^\frac72}{u_\Lambda^2}\left(
\dot y^2 +h(y^2-2zy\dot y)\right)\right]
\ee
where a dot denotes derivative with respect to $z$ and
$\tilde T=T_4 e^{-\phi}\frac25\frac{u_\Lambda^{3/2}}{R_{AdS}^3}$.
Finally, in order to simplify this expression, let us define:
\be
u_z(z)=(u_\Lambda^5 + u_\Lambda^3 z^2)^\frac15
\ee
and subtract a total derivative $\partial_z \left(u_z^\frac72
h\,z\, y^2/(2 u_\Lambda^2)\right)$ so we find:
\be
S_{D4}=-\tilde T \int d^4x\,dz\left[
\frac{2}{25}\,\frac{R_{AdS}^4 u_\Lambda}{u_z^{7/2}}\eta^{\mu\nu}\partial_{\mu} y\,\partial_{\nu} y
+\frac{7}{10}\frac{u_\Lambda}{u_z^{3/2}}y^2+ \frac12 
\frac{u_z^{7/2}}{u_\Lambda^2} \dot y^2 \right]\,\,.
\label{yfluct}
\ee
We want to obtain the spectrum of normalizable small fluctuations
associated to the action (\ref{yfluct}). From the four-dimensional 
point of view, they  correspond to (pseudo) scalar excitations.
It will be useful to define several quantities:
\be
Z\equiv\frac{z}{u_\Lambda}\,\,,\qquad\qquad
K\equiv\left(\frac{u_z}{u_\Lambda}\right)^5=1+Z^2\,\,,
\ee
In terms of these quantities, the action for the fluctuation reads:
\be
S_{D4}=-\frac{4}{25}\tilde T \frac{R_{AdS}^4}{u_\Lambda^\frac32}
\int d^4x dZ \left[ \frac12 K^{-\frac{7}{10}} \eta^{\mu\nu}\partial_{\mu} y\,\partial_{\nu} y+
\frac{M_\Lambda^2}{2}\left(\frac75 K^{-\frac{3}{10}}\, y^2
+K^{\frac{7}{10}} (\partial_Z y)^2\right)\right]
\ee
Now, by expanding:
\be
y(x^\mu, Z) =\sum_{n=1}^\infty {\cal U}^{(n)}(x^\mu) \rho_{(n)} (Z)
\ee
the equation for the set of $\rho_{(n)}$ reads:
\be
-K^{\frac{7}{10}}\,\partial_Z (K^{\frac{7}{10}} \partial_Z \rho_{(n)})
+\frac75 K^\frac{4}{10} \rho_{(n)} = \left(\frac{m'_n}{M_\Lambda}\right)^2
 \rho_{(n)}\,\,,
\label{scalareq2}
\ee
and the orthonormality condition is:
\be
\frac{4}{25}\tilde T \frac{R_{AdS}^4}{u_\Lambda^\frac32}
\int dZ K^{-\frac{7}{10}}\rho_{(n)} \rho_{(m)} =\delta_{mn}
\label{scalarnorm2}
\ee
Notice that from equation (\ref{scalareq2}), it is easy to find out
the behavior of the $\rho_{(n)}$ at infinity:
\be
\lim_{Z\to \infty} \rho_{(n)} \sim C_1 Z + \frac{C_2}{Z^\frac75}
\ee
From (\ref{scalarnorm2}) we see that normalizability of the mode requires
$C_1=0$ thus yielding a discrete spectrum.
 Since (\ref{scalareq2}) is invariant under $Z \to -Z$
the solutions have well defined (odd or even) parity under this
transformation, leading to mesons of different parity and charge
conjugation as discussed in \cite{SS}. There are, therefore, 
two possible boundary conditions at $Z=0$:
\ba
\partial_Z \rho_{(n)} (0) =0 \quad &\rightarrow&
0^{++}\qquad \textrm {(odd $n$)}\nonumber\\
\rho_{(n)} (0) =0 \quad &\rightarrow&
0^{--}\qquad \textrm {(even $n$)}
\ea
The values obtained numerically by inserting these conditions on
(\ref{scalareq2}) and requiring normalizability
 are reported on the main text (\ref{scmasses}).


\newsection{Spinning open string calculation details}\label{app: spinning}

In this appendix we report the calculations of the spectrum of large angular momentum mesons, evaluated form the dual  non-critical background. The calculation is very similar  to what has been done in some critical examples \cite{KMMW, KPSV}.

We start from a set of coordinates for the transverse space to the D4 color branes, which is closely related to the one introduced in (\ref{newcoords}). Let us take
\be
u^5=u^5_\Lambda(1+\rho^2) \qquad \quad \tilde{\eta}=\frac{5 u_\Lambda}{2R^2}\eta
\ee
with respect to which the metric reads
\be\label{ds2bis}
ds^2=\!\left(\frac{u_\Lambda}{R_{AdS}}\right)^2(1+\rho^2)^{\frac{2}{5}}(-(dx^0)^2+(dx^3)^2+dR^2+R^2\,d\phi^2)+\frac{4}{25}R_{AdS}^2\left(\frac{d\rho^2}{1+\rho^2}\!+\!\frac{\rho^2\,d\tilde{\eta}^2}{(1+\rho^2)^\frac{3}{5}}\right)
\ee

Since we consider an open string rotating on the $(x^1,x^2)$ plane, that is along the angle~$\phi$, we take the following ansatz for the motion of the string
\begin{align}\label{spineq}
&X^0=\tau & &\phi=\omega \tau & &X^3=const.\\
& R(\sigma,\tau)=R(\sigma)& &\rho(\sigma,\tau)=\rho(\sigma) & &\tilde{\eta}(\sigma,\tau)=\tilde{\eta}(\sigma)
\end{align}
Moreover, the endpoints of the string need to be attached to the probe D4, that is
\be
\rho( {\textstyle \pm \frac{\pi}{2}} )=\rho(u_{D4}(\eta))
\ee
where $u_{D4}(\eta)$ is the profile of the probe D4-brane (\ref{embed}).
The Nambu-Goto action for this configuration reads then
\be\label{osact}
\begin{split}
S_{NG}=&\frac{1}{2\pi\alppr}\int_\Sigma d^2\sigma \sqrt{-h_{\alpha \beta}}=\\
 =&T_q \int_\Sigma d^2\sigma (1+\rho^2)^\frac{2}{5}\sqrt{((\dot{X}^0)^2)-R^2\dot{\phi}^2)(R'^2+\frac{4}{25}\frac{R_{AdS}^4}{u_\Lambda^2}(\frac{\rho'^2}{(1+\rho^2)^\frac{7}{5}}+\frac{\rho^2}{1+\rho^2}\tilde{\eta}'^2))}
\end{split}
\ee 
where $h_{\alpha\beta}$ is the string worldsheet metric induced by the background (\ref{ds2bis}), and \be
T_q\equiv \frac{1}{2\pi\alppr}\left(\frac{u_\Lambda}{R_{AdS}}\right)^2
\ee
is the effective tension for a string stretching close to the horizon of the $AdS_6$ black hole (\ref{ds2bis}) \cite{KPSV}.

The action (\ref{osact}) is independent of $X^0$ and $\phi$, therefore the following  quantities are conserved
\begin{align}
&E=T_q\int  
d\sigma (1+\rho^2)^\frac{2}{5}\sqrt{\frac{R'^2+\frac{4}{25}
\frac{R_{AdS}^4}{u_\Lambda^2} \left(\frac{\rho'^2}{(1+\rho^2)^{7/5}}
+\frac{\rho^2}{1+\rho^2}\tilde{\eta}'^2 \right) }{1-\omega^2R^2}} 
\label{E}\\[7pt]
&J=T_q \,
\omega \int  d\sigma R^2(1+\rho^2)^\frac{2}{5}\sqrt{\frac{R'^2+
\frac{4}{25}\frac{R_{AdS}^4}{u_\Lambda^2} \left(\frac{\rho'^2}
{(1+\rho^2)^{7/5}}+\frac{\rho^2}{1+\rho^2}\tilde{\eta}'^2 \right) }
{1-\omega^2R^2}} \label{J}
\end{align}

The variation of the action with respect to $\tilde{\eta}'$ is given by
\be
\left.\delta S_{NG}\right|_{\delta\tilde{\eta}'}=\frac{4R_{AdS}^2}
{50\pi\alppr}\int_\Sigma d^2\sigma \frac{\rho^2}{(1+\rho^2)^\frac{3}{5}} 
\sqrt{\frac{1-\omega^2R^2}{R'^2+\frac{4}{25}\frac{R_{AdS}^4}{u_\Lambda^2}
\left(\frac{\rho'^2}{(1+\rho^2)^{7/5}}+\frac{\rho^2}{1+\rho^2}\tilde{\eta}'^2
\right)}}\;\tilde{\eta}'\delta\tilde{\eta}'
\ee
and since the action does not depend on $\tilde{\eta}$, for the above variation to vanish it must be either $\tilde{\eta}'=0$ or $\rho=0$. For $u_0>u_\Lambda$ only the first condition can be satisfied, and moreover setting $\tilde{\eta}'=0$ always lowers the energy, as can be easily seen from (\ref{E}). Therefore from now on we take $\tilde{\eta}$ to be constant for the embedding of the open string.

The total variation of the open string action (\ref{osact}) reads, after integrating by parts,
\be\label{SNGvar}
\begin{split}
\delta S_{NG}=& T_q\int_\Sigma d^2\sigma  \left\{ -\left[\sqrt{
\frac{R'^2+\frac{4}{25}\frac{R_{AdS}^4}{u_\Lambda^2}\frac{\rho'^2}
{(1+\rho^2)^{7/5}}}{1-\omega^2R^2}} (1+\rho^2)^\frac{2}{5}\omega^2  R  
 +\right.\right.\\
&+\left.\partial_\sigma \left(  \sqrt{\frac{1-\omega^2R^2}
{R'^2+\frac{4}{25}\frac{R_{AdS}^4}{u_\Lambda^2}\frac{\rho'^2}
{(1+\rho^2)^{7/5}} }    }    (1+\rho^2)^\frac{2}{5} R'\right)
\right]\delta R+\\
&+\frac{4}{5}\left[ \sqrt{\frac{1-\omega^2R^2}{R'^2+\frac{4}{25}
\frac{R_{AdS}^4}{u_\Lambda^2}\frac{\rho'^2}{(1+\rho^2)^{7/5}} }    } 
 \frac{\rho}{(1+\rho^2)^{3/5}}\left( R'^2-\frac{3}{25}\frac{R_{AdS}^4}
 {u_\Lambda^2} \frac{\rho'^2}{(1+\rho^2)^{7/5}}\right)  \right.+\\
&-\left.\left.\partial_\sigma \left(   \sqrt{\frac{1-\omega^2R^2}
{R'^2+\frac{4}{25}\frac{R_{AdS}^4}{u_\Lambda^2}\frac{\rho'^2}
{(1+\rho^2)^{7/5}} }    }        \frac{R_{AdS}^4}{5u_\Lambda^2}
\frac{\rho'}{1+\rho^2}\right)\right]\delta\rho\right\}+\\
 &+T_q\int 
d\tau\! \left. \sqrt{\frac{1-\omega^2R^2}{R'^2+\frac{4}{25}\frac{R_{AdS}^4}
{u_\Lambda^2}\frac{\rho'^2}{(1+\rho^2)^{7/5}} }    }    
(1+\rho^2)^\frac{2}{5}\left(   R' \delta R+\frac{4}{25}
\frac{R_{AdS}^4}{u_\Lambda^2}\frac{\rho'\delta\rho}{(1+\rho^2)^{7/5}} 
\right)\right|_{\sigma=-\frac{\pi}{2}}^{\sigma=\frac{\pi}{2}}
\end{split}
\ee
The boundary term in the last line fixes the boundary conditions for the endpoints of the open string. There are two ways to set this term to zero
\be\label{boundconds}
\omega R\left(\pm\frac{\pi}{2}\right)=1\qquad \mathrm{or}\qquad \left\{  \begin{array}{l}  \left.R'\,\delta R \right|_{\sigma=\pm\frac{\pi}{2}} =0   \\       \left.\rho'\,\delta \rho \right|_{\sigma=\pm\frac{\pi}{2}}     =0\end{array}  \right.
\ee
It has been  shown in \cite{KMMW} that for a configuration similar to ours, imposing that the string endpoints move at the velocity of light does not allow for both of them to be attached to the same brane. Therefore the first condition in~(\ref{boundconds}) has to be discarded. 
We are only left, then, with  the second set of boundary conditions. For the first condition in this set, we notice that the probe brane lies along the $R$ direction, and therefore it is natural to impose $R'=0$ while leaving $\delta R$ arbitrary: the string attaches perpendicularly to the brane, but its endpoints can move freely over it. For the last condition,  it is sufficient to notice that the minimal energy condition requires the endpoints of the string to be as close as possible to the horizon of the black hole, that is the string needs to be attached to the lowest point of the probe brane
\be
\rho\left(\pm\frac{\pi}{2}\right)=\rho(u_0)\equiv \rho_0
\ee
and therefore
\be
\delta\rho\left(\pm\frac{\pi}{2}\right)=0
\ee
which satisfies the remaining boundary condition (\ref{boundconds}).

There are two possible convenient choices to gauge the $\sigma$ parameterization invariance of the string worldsheet, either $R=\sigma$ or $\rho=\sigma$. In the first case, the equation of motion coming from setting (\ref{SNGvar}) to zero reads
\be
\begin{split}
\frac{\rho''}{\rho'}-\frac{11}{5}\frac{\rho'\rho}{1+\rho^2}-5\frac{u_\Lambda^2}{R_{AdS}^4}(1+\rho^2)^\frac{2}{5}\frac{\rho}{\rho'}-\frac{\omega^2R}{1-\omega^2R^2}\left( 1+\frac{4}{25}\frac{R_{AdS}^4}{u_\Lambda^2}\frac{\rho'^2}{(1+\rho^2)^{7/5}}  \right)=0
\end{split}
\label{rhoeq}
\ee
while fixing $\rho=\sigma$ one obtains
\be
\frac{R''}{R'}+\frac{11}{5}\frac{\rho}{1+\rho^2}
+5\frac{u_\Lambda^2}{R_{AdS}^4}\rho(1+\rho^2)^\frac{2}{5}R'^2+
\frac{\omega^2 R}{R'(1-\omega^2 R^2)}
\left(R'^2+\frac{4}{25}\frac{R_{AdS}^4}
{u_\Lambda^2}\frac{1}{(1+\rho^2)^{7/5}}\right)
=0
\ee

{\it \noindent  Rotating long string analysis}

These equations are quite involved and most probably finding an exact solution might prove impossible. Nonetheless a numerical solution can certainly be found, by imposing appropriate boundary conditions. For the moment let us focus, though, on an interesting limit. When the distance $2R_0$ between the open string endpoints is very large, the angular velocity needs to be really small (for fixed quark masses) and the string configuration resembles very much the one encountered in the Wilson loop calculation. To a fairly good extent, therefore, it can be approximated by a piecewise-straight string with two sides stretching along the $\rho$ direction from the probe brane to the black hole horizon at $\rho=0$ and a third piece lying along the $R$ direction at $\rho=0$. This configuration has to be supplemented by an additional set of boundary conditions, requiring that the two internal vertices of the string (we can always choose a parameterization in which they are at  $\sigma=\pm \alpha$) connecting the ``vertical'' and ``horizontal'' straight pieces,  have to lie on the horizon of the black hole
\be
\rho(\pm \alpha)=0\quad\Rightarrow\quad \delta\rho|_{\sigma=\pm\alpha}=0
\ee

By substituting this simplifying ansatz in the Nambu-Goto action variation (\ref{SNGvar}), the only surviving terms are
\be
\begin{split}
\delta S_{NG}=T_q&\left\{-\frac{2}{5}\frac{R_{AdS}^2}{u_\Lambda}\frac{\omega^2 R_0}{\sqrt{1-\omega^2 R_0^2}} \int d\tau \left(\int_{-\frac{\pi}{2}}^{-\alpha} d\sigma+\int^{\frac{\pi}{2}}_{\alpha} d\sigma\right)\left(  \frac{\rho'}{(1+\rho^2)^{3/10}} \right)\delta R+\right.\\
&+\left.  2 \sqrt{1-\omega^2 R_0^2} \int d\tau \delta R \right\}
\end{split}
\ee
and stationarity of the action can be achieved only by imposing $\delta R(\sigma,\tau)=\delta R(\tau)$ and enforcing the relation
\be\label{eq param}
T_q(1-\omega^2 R_0^2)=\omega^2 R_0 m_q
\ee
where, in these coordinates, the definition (\ref{quarkmass}) reads:
\be
m_q\equiv\frac{1}{2\pi \alppr} \int_0^{\rho_0} d\rho \sqrt{-g_{00}g_{\rho\rho}}=\frac{1}{2\pi\alppr}\frac{2u_\Lambda}{5}\int_0^{\rho_0}\frac{d\rho}{(1+\rho^2)^{3/10}}
\ee
Apart from the dependence of the effective string tension $T_q$ on the ratio $u_\Lambda/R_{AdS}$, equation (\ref{eq param}) is the same as the relation obtained in \cite{KPSV}. Equation (\ref{eq param}) can be easily solved for $R_0$
\be\label{R0}
R_0=\frac{m_q}{2T_q}\left( \sqrt{1+\frac{4T_q^2}{m_q^2 \omega^2}} -1 \right)
\ee

We can now evaluate the energy and angular momentum of the string configuration in the limit of large $R_0$. We find
\begin{align}
&E=\frac{2T_q}{\omega}\left(  \frac{\sqrt{1-\omega^2 R_0^2}}{\omega R_0} +\arcsin  (\omega R_0) \right)\label{Esos}\\
& J=\frac{T_q}{\omega^2}\left(  \omega R_0  \sqrt{1-\omega^2 R_0^2} +\arcsin  (\omega R_0) \right)\label{Jsos}
\end{align}
where the interquark distance $2R_0$ is determined in terms of the quarks' mass and angular velocity $\omega$ through (\ref{R0}). These relations are solidly reliable only in the large $J$ limit which, for fixed $m_q$, is the one corresponding to the small $\omega$ region. In this regime it is easy to show that $R_0\simeq 1/\omega$ and the energy and angular momentum of the string configuration satisfy a Regge law
\be
E^2=2\pi T_q J
\ee

{\it \noindent Rotating short strings analysis}

We analyze here the limit in which $\omega \to \infty$ and therefore
$R_0 \to 0$. Since the string is very short, it does not really get affected
by the particular background where it is embedded and we will recover
a result very similar to that found in \cite{KMMW}.

Let us consider:
\be
\rho = \rho_0 + \delta\rho (R)
\ee
Keeping only leading order terms\footnote{We are assuming
$1\gg \frac{R_{AdS}^4}{u_\Lambda^2}\,\frac{\rho'^2}{(1+\rho^2)^7/5}$.
This is true except near the endpoints where of course
$\rho'\to \infty$. What happens here is analogous to 
\cite{KMMW}.}
 in $\delta\rho$ we find from
(\ref{rhoeq}):
\be
\delta\rho'' - K - \frac{\omega^2 R }{1-\omega^2 R^2}\delta\rho'=0
\ee
where the constant $K$ has been defined as:
\be
K=5\frac{u_\Lambda^2}{R_{AdS}^4} (1+\rho_0)^{2/5} \rho_0
\ee
This equation of motion has to be supplemented with the
initial condition $\delta\rho'(0)=0$, coming from the 
$R \to -R$ symmetry of the problem. It is immediate to find:
\be
\delta\rho' = \frac{KR}{2}+\frac{\arcsin(R\omega)}{2\omega \sqrt{1-R^2\omega^2}}
\ee
Since we need $\delta\rho'(R_0)=\infty$, we have $R_0 \to \omega^{-1}$.
Substituting in (\ref{E}) and (\ref{J}) one finds that at leading
$\omega \to \infty$ order,
$E\propto \omega^{-1}$ and $J\propto \omega^{-2}$ so there is again a 
Regge behavior
\be
E^2 = 2 \pi \tau_{eff} J
\ee
with:
\be
\tau_{eff} = \frac{1}{2\pi \alpha'} \left(\frac{u_\Lambda}{R_{AdS}}\right)^2
(1+\rho_0^2)^{2/5}
\ee
Notice that $\tau_{eff} > T_q$.



\end{document}